\documentclass[aip,reprint,superscriptaddress,showkeys]{revtex4-1}

\usepackage[a4paper,left=1in,top=1in,right=1in,bottom=1in]{geometry}

\usepackage{epsf,graphicx,amsfonts,amsmath,amssymb}
\usepackage{soul}
\usepackage[latin1]{inputenc}
\usepackage[final]{changes}
\usepackage{array}
\usepackage{multirow}
\usepackage{graphicx}

\font\bbfnt=msbm10

\def\bbR{\mbox{\bbfnt R}}

\newcommand{\mb}[1]{\mbox{\bfseries \itshape #1}}
\newcommand{\1}{\Bar{1}}
\newcommand{\2}{\Bar{\Bar{1}}}
\newtheorem{defi}{Definition}

\definechangesauthor[color=magenta]{ClL}
\definechangesauthor[color=blue]{ChL}
\definechangesauthor[color=orange]{ChG}

\usepackage{color}

\usepackage{supertabular}

\begin{document}

\title{Assessing observability of chaotic systems using Delay Differential 
Analysis}

\author{
Christopher E. Gonzalez$^{*,1,2}$,
Claudia Lainscsek$^{*,1,3}$,
Terrence J. Sejnowski$^{1,3,4}$,
and
Christophe Letellier$^{5}$\\}

\affiliation{
$^*$These authors contributed equally to this work.\\
$^.$Corresponding author: Christopher Gonzalez cegonzalez@ucsd.edu \\
$^1$Computational Neurobiology Laboratory, The Salk Institute for Biological Studies, 10010 North Torrey Pines Road,
La Jolla, CA 92037, USA \\
$^2$ Department of Neurosciences, University of California San Diego, La Jolla, California 92093, USA \\
$^3$ Institute for Neural Computation, University of California San Diego, La Jolla, CA 92093, USA \\
$^4$ Division of Biological Sciences, University of California San Diego, La Jolla, CA 92093, USA \\
$^5$Rouen Normandie Universit\'e, CORIA, Campus Universitaire du 
Madrillet, F-76800 Saint-Etienne du Rouvray, France
}

\begin{abstract}
Observability can determine which recorded variables of a given system are 
optimal for discriminating its different states. Quantifying observability 
requires knowledge of the equations governing the dynamics. These equations 
are often unknown when experimental data are considered. Consequently, we 
propose an approach for numerically assessing observability using Delay 
Differential Analysis (DDA). Given a time series, DDA uses a delay differential 
equation for approximating the measured data.
The lower the least squares error between the predicted and recorded 
data, the higher the observability. We thus rank the variables of 
several chaotic systems according to their corresponding 
least square error to assess observability. 
The performance of our approach is evaluated by comparison with the ranking 
provided by the symbolic observability coefficients as well as with two 
other data-based approaches using reservoir computing and singular value 
decomposition of the reconstructed space. We investigate the 
robustness of our approach against noise contamination. 
\end{abstract}

\date{\today}

\maketitle

{\bf  
A popular approach for studying nonlinear dynamical systems from a recorded 
time series is to reconstruct the original system using delay or derivative 
coordinates. It is known that the choice of the measured variable can affect 
the quality of attractor reconstruction. Unlike in linear systems for 
which the state space is observable or not from the measurements, nonlinear 
systems are more or less observable from measurements depending on the state 
space location. Moreover the observability strongly depends
on the measured variables. It is therefore useful to assess the observability 
provided by a variable using a real number within the unit interval between two 
extreme values: 0 for nonobservable, and 1 for full observability. Analytical 
techniques for determining observability require knowledge of the 
underlying equations which are typically unknown when an experimental system is 
investigated. This is often the case for social and biological networks.
It is thus of primary importance to assess observability directly 
from recorded time series. In this paper, we show how Delay Differential 
Analysis (DDA) can assess observability from time series. The 
performance of this approach is evaluated by comparing our results obtained for 
simulated chaotic systems with the symbolic observability coefficients obtained 
from the governing equations. 
}

\section{Introduction}

Studying dynamical systems from real world data can be difficult as they are 
often high-dimensional and nonlinear; moreover, it is typically not possible to 
measure all the variables spanning the associated state 
space.\cite{Liu13,Wan14,Wha15,Lei17,Hab18,Lei18,Let18} In theory, 
it is possible to reconstruct the non-measured variables by using delay or 
differential embeddings from a single measurement.\cite{Tak81} However, when 
performing state-space reconstruction, the dimension required to obtain a 
diffeomorphical equivalence --- required for correctly distinguishing the 
different states of the system --- with the original state space may depend on 
the 
measured variable(s).\cite{Let98b} Indeed, a $d$-dimensional system 
can be optimally reconstructed from a given variable with a $d$-dimensional 
embedding but a higher-dimensional space may be required when another variable
is measured. For instance, the R\"ossler attractor is easily reproduced with a 
three-dimensional global model from variable $y$ but a four-dimensional 
model\cite{Let98b} or a quite sophisticated procedure\cite{Lai03} is
needed when variable $z$ is measured. It was shown that data analysis often (if not always) depends on the observability provided by the measured
variable.\cite{Let06c,Let06b,Por19} 

In the 1960s, the concept of observability was introduced by Rudolf 
K\'{a}lm\'{a}n in control theory.\cite{Kal60} Observability assesses 
whether different states of the original system can be distinguished from the 
measured variable. A system is said to be 
fully observable from some measurement if the rank of the observability matrix 
is equal to the dimension of the system.\cite{Kai80,Che99} 
With such an approach, 
the answer is either fully observable or non observable. This approach is 
sufficient for linear systems because the observability matrix does not depend 
on the location in the state space. 

This is not true for nonlinear systems and 
observability coefficients were introduced to overcome this inaccurate answer.\cite{Agu95,Let98b} Observability coefficients are real 
numbers within the unit interval between two extreme values, 0 for 
nonobservable, 1 for fully observable. These coefficients are estimated at 
every point of the trajectory produced by the governing equations in the state
space, and then averaged along that trajectory.\cite{Agu95,Let98b} It
is also possible to construct symbolic observability coefficients from the 
Jacobian matrix of the system studied.\cite{Let09,Bia15} In this way, 
observability takes a graded value according to the probability with which the 
attractor intersects the singular observability 
manifold,\cite{Fru12} that is, the subset of the original space for 
which the determinant of the observability matrix is zero.
The great advantage of these coefficients is that they allow comparing the 
observability provided by variables from different systems and they can be 
computed for high-dimensional systems.\cite{Let18} It is then possible to rank 
the variables according to the observability of the original state space they 
provide. The dependency of the observability on
the measured variable is due to the way variables are coupled in the original 
system.\cite{Let05b} Symmetries are often sources of difficulty for assessing 
observability, particularly because reconstructing the original symmetry is not 
possible from a single variable if the symmetry differs from an 
inversion.\cite{Kin92,Let02} 

The weakness of these analytical approaches is that the governing equations 
must be known and it is not possible to assess observability from experimental 
data. A first attempt to overcome this was based on a singular value 
decomposition of some matrices built from local data.\cite{Agu11} 
Results were encouraging but some slight discrepancies with analytical results 
were noticed. Another approach, based on a model built directly from the
data using reservoir computing was also proposed.\cite{Car18} In both cases,
some discpreancies with the symbolic observability coefficients were observed. 
It therefore remains challenging to develop a reliable technique which always 
matches with theoretical results. In this work we propose a measure for 
assessing observability from recorded data by using delay differential analysis 
(DDA) and compare our results and those obtained --- when available in 
the literature --- with the two techniques 
discussed above with the symbolic observability coefficients computed
for several well-studied chaotic systems. Here, DDA is based on a delay 
differential equation which approximates the dynamics underlying the measured 
time series. Contrary to what is done with global modeling\cite{Agu09} or 
reservoir computing,\cite{Pat17} there is no need for an accurate model. Previous work showed a rough model with a very limited number of terms 
(typically three) is sufficient to detect dynamical changes or
classify various dynamical regime. \cite{Lai00,Lai13,Lai15c} 

The subsequent part of this paper is organized as follows. 
Section \ref{sym_ob} is a brief introduction to the computation of symbolic 
observability coefficients. 
Section \ref{DDAn}
provides an introduction to DDA and explains how it can be used for ranking
variables according to the observability of the state space they provide. 
Section \ref{resu} introduces the investigated chaotic systems and 
provides the corresponding symbolic observability coefficients. Section
\ref{DDAran} is the main section of this 
paper: it discusses the performance of DDA for assessing observability of 
the chaotic systems and compares it with those
of the two other data-based techniques. Section \ref{conc} provide 
some conclusions.

\section{Theoretical Background}

\subsection{Symbolic observability coefficients}
\label{sym_ob}

Let us consider a $d$-dimensional dynamical 
system represented by the state vector $\mb{x} \in \bbR^d$ whose components are 
given by
\begin{equation}
  \label{system}
  \dot{x}_i = f_i (x_1,x_2,x_3, ..., x_d), ~~~~~ i=1,2,3,...,d
\end{equation}
where $f_i$ is the $i$th component of the vector field $\mb{f}$. Let us 
introduce the measurement function $h(\mb{x}): \mathbb{R}^d \mapsto 
\mathbb{R}^m$ of $m$ variables chosen among the $d$ ones spanning the original 
state space. It is then required to reconstruct a space 
$\mathbb{R}^{d_{\rm r}}$ ($d_{\rm r} \geq d$) from the $m$ measured variables. 
One has to choose $d_{\rm r}-m$ derivatives of these $m$ measured 
variables to get a $d_{\rm r}$-dimensional vector $\mb{X}$ spanning the 
reconstructed space. Commonly, observability is assessed by using 
$d_{\rm r} = d$.\cite{Kai80,Che99} In the present work, we are only 
working 
with scalar time series ($m=1$). The change of coordinate between the original
state space and the reconstructed one is thus the map
\begin{equation}
  \Phi: \mathbb{R}^d (\mb{x}) \mapsto \mathbb{R}^d (\mb{X}) \, . 
\end{equation}
When the derivative coordinates are used for spanning the reconstructed
space, the map can be analytically computed.\cite{Let05a}
The observability of a system from a variable is defined as 
follows.\cite{Kai80,Chen99} For the sake of simplicity, let us limit ourselves 
to the case $m=1$ (a generalization to the others cases is straightforward). 
\begin{defi}
The dynamical system (\ref{system}) is said to be {\it state observable} at 
time $t_f$ if every initial state $\mb{x} (0)$ can be uniquely determined from 
the knowledge of the vector $s (\tau)$, $ 0 \leq \tau \leq t_f$. 
\end{defi}
To test whether a system is observable or not is to construct
the observability matrix\cite{Her77} which is defined as the Jacobian 
matrix of the 
Lie derivatives of $h(\mb{x})$. Differentiating $h(\mb{x})$ yields
\[ \frac{{\rm d}}{{\rm d}t}h(\mb x) 
   =\frac{\partial h}{\partial \mb{x}}\dot{\mb x} 
   =\frac{\partial h}{\partial \mb{x}}\mb{f}(\mb{x})
   ={\cal L}_f h(\mb{x}) \, ,
\]
where ${\cal L}_fh(\mb x)$ is the Lie derivative of $h (\mb{x})$ along the 
vector field $\mb f$. The $k$th order Lie derivative is given by
\[ {\cal L}_{\mb{f}}^kh(\mb x)
   = \frac{\partial {\cal L}_{\mb{f}}^{k-1} h (\mb{x})}
          {\partial \mb x}{\mb f}(\mb x) \, ,
\]
being the zero order Lie derivative the measured variable itself,
${\cal L}_f^0h(\mb x)=h(\mb x)$. Therefore, the observability matrix
${\cal O} \in \mathbb{R}^{d\times d}$ is written as
\begin{equation}
  {\cal O} (\mb x) = 
  \left[
    \begin{array}{l}
      {\rm d} h(\mb{x}) \\[0.1cm]
      {\rm d} {\cal L}_f h(\mb{x}) \\[0.1cm]
      \vdots \\[0.1cm]
      {\rm d} {\cal L}^{d-1}_f h(\mb{x}) 
    \end{array}
  \right]
\end{equation}
where ${\rm d}\equiv \frac{\partial}{\partial \mb x}$. The observability 

\begin{defi}
The dynamical system (\ref{system}) is said to be state observable if and only 
if the observability matrix has full rank, that is, rank$({\cal O})=d$. 
\end{defi}
The observability matrix ${\cal O}$ is equal to the Jacobian matrix of the 
change of coordinates $\Phi: \mb{x} \rightarrow \mb{X}$ when derivative
coordinates are used.\cite{Let05a} In this approach, the observability 
is either full or zero. The term {\it structural} was introduced when the 
results do not depend on parameter values\cite{Lin74}. Computing the rank of 
the observability matrix is independent of parameter values and, consequently, 
is an example of structural observability.\cite{Agu18}
Computing observability with graphs\cite{Lin74,Liu13,Let18b} is also a structural
approach. We term observability assessed from recorded data --- necessarily
dependent on the parameter values used for simulating the trajectory of the 
system ---as {\it dynamical} observability.\cite{Agu18} This
type of approach returns a real number within the unit interval: variables can
be ranked between the two extreme cases, 1.0 (0.0) for a full (null) 
observability 
There is a third type of observability,  {\it symbolic} 
observability, which does not depend on parameter values but allows ranking
the variables.\cite{Bia15} All types of observability are not sensitive to symmetry-related problems. This is 
due to the fact that observability is a local property while symmetry is a global one. Consequently, symmetry may
degrade the assessment of observability.\cite{Let02}

The procedure to compute symbolic observability coefficients is implemented in 
three steps as follows.\cite{Bia15,Let18} First, the Jacobian 
matrix ${\cal J}$ of the system (\ref{system}), composed of elements $J_{ij}$, 
is transformed into the symbolic Jacobian matrix $\tilde{\cal J}$ by replacing 
each constant element $J_{ij}$ by 1, each polynomial element $J_{ij}$ by $\1$, 
and each rational element $J_{ij}$ by $\2$ when the $j$th variable is present 
in the denominator, or by $\1$ otherwise. Rational terms in the governing 
equations (\ref{system}) are distinguished from polynomial terms since the 
formers reduce more strongly the observability than the latters.\cite{Bia15}

Then the symbolic observability matrix $\tilde{\cal O}$ is constructed. 
The first row of $\tilde{\cal O}$ 
is defined by the derivative of the measurement function ${\rm d} h (\mb x)$, 
that is, $\tilde O_{1j}=1$ if $j=i$ and $0$ otherwise when the $i$th variable 
is measured. The second row is the $i$th row of $\tilde{\cal J}$. The $k$th row 
is obtained as follows. First, each element of the $i$th row of 
$\tilde {\cal J}$ is multiplied by the corresponding $i$th component of the 
vector $\mb v=(\tilde{O}_{\ell 1},\cdots,\tilde{O}_{\ell d})^{\rm T}$ where 
$\ell =k-1$ refers to the $(k-1)$th row of the symbolic observability matrix 
$\tilde{\cal O}$. The rules to perform the symbolic product 
$\tilde{ J}_{ij} \otimes  v_i$ are such that \cite{Bia15}
\begin{equation}
  \left|
    \begin{array}{ll}
      0 \otimes a & = 0, \\[0.1cm]
      1 \otimes a & = a, \\[0.1cm]
      \1 \otimes a & = a \text{ for } a=\1,\2,\\[0.1cm]
      \2 \otimes a & = \2 \text{  for } a\ne 0.
    \end{array}
  \right.\label{otimes}
\end{equation}
Second, the matrix $\tilde{\cal J}'$ is reduced into a row where each element 
$\tilde{ O}_{kj}=\sum_i \tilde{ J}'_{ij}$ according to the addition 
law\cite{Bia15}
\begin{equation}
  \left|
    \begin{array}{ll}
      0 \oplus a & = a, \\[0.1cm]
      1 \oplus a & = a \text{ for } a\ne 0,\\[0.1cm]
      \1 \oplus a & = a \text{ for } a\ne 0,1,\\[0.1cm]
      \2 \oplus a & = \2.
    \end{array}
  \right.
\end{equation}

The last step is associated with the computation of the symbolic observability 
coefficients. The determinant of $\tilde{\cal O}$ is computed according to the 
symbolic product rule defined in~(\ref{otimes}) and expressed as products and 
addends of the symbolic terms $1$, $\1$ and $\2$, whose number of occurrences 
are $N_1$, $N_{\1}$ and $N_{\2}$, respectively. It is
convenient to impose that, if $N_{\1}=0$ and $N_{\2}\neq 0 $ then 
$N_{\1}=N_{\2}$. The symbolic observability coefficient is thus defined as
\begin{equation}
  \eta = \displaystyle 
         \frac{1}{D}{N_1} +\frac{1}{D^2}N_{\1}+\frac{1}{D^3} N_{\2}  
\end{equation}
with $D= N_1 + N_{\1} + N_{\2}$. This coefficient is in the unit 
interval, $\eta =1$ for a variable providing full observability of the original
state space. An observability is said to be good when 
$\eta \geq 0.75$.\cite{Sen16}

\subsection{Delay Differential Analysis}
\label{DDAn}

Let us assume that a time series $\left\{ X_1\right\}$ is recorded in a 
$d$-dimensional system. From this time series, it is possible to obtain a 
global model reproducing the underlying dynamics. There are typically two main
approaches working with either derivative or delay 
coordinates.\cite{Cru87,Agu09} When derivatives are used, it is 
possible to construct a $d$-dimensional 
differential model
\begin{equation}
  \label{glomod}
  \left\{
    \begin{array}{l}
      \dot{X}_1 = X_2 \\[0.1cm]
      \dot{X}_2 = X_3 \\[0.1cm]
      \vdots \\
      \dot{X}_d = F \left( \displaystyle X_1, X_2, ..., X_d \right) 
    \end{array}
  \right.
\end{equation}
where $X_i$ is the $(i-1)$th derivative of the measured variable 
$X_1$.\cite{Gou94} The function 
$F$ can be numerically estimated by using a least-square technique with a 
structure selection.\cite{Man12,Man19} $F$ can be polynomial\cite{Gou94,Man19} 
or rational.\cite{Lai01,Lai11} This model requires $d$-ordinary differential 
equations whose variables are the $d$ successive derivatives of $X_1$: this model 
works in a differentiable embedding.

Second, it is possible to construct a model whose equations have the form of a
difference equation
\begin{equation}
  \label{diffeq}
	X (k+1) = F \left( \displaystyle X_{\tau_j} (k) \right)
	= \sum_{i=0}^N{\displaystyle a_i \, \varphi_i 
	}
\end{equation}
where $\varphi_i$ is a monomial of delay coordinates 
$X_{\tau_j} (k) = X (k-\tau_j)$ with $\tau_j = n \delta_t$ ($n \in 
\mathbb{N}^+$) is a time delay expressed in terms of the sampling time 
$\delta_t$ with which the scalar time series $\{ X_1(k) \}$
is recorded: $k$ is the discrete time.
Such a model has an auto-regressive form, and typically the number $N$ of terms 
is between 10 and 20. The space in which this model is working is thus spanned
by delay coordinates: its 
dimension is
very often significantly larger than the dimension $d$, the embedding 
dimension\cite{Cao97a} or even than the Takens criterion.\cite{Tak81} An optimal
form of the difference equation (\ref{diffeq}) is developed under the form of a
nonlinear autoregressive-moving average (NARMA) model.\cite{Agu95a}

Recently a third type of model was investigated under the name of {\it 
reservoir computing}.\cite{Luk09} This approach considers an oversized 
model with a functional structure based on a network whose nodes are 
characterized by some simple function. For instance, the Lorenz attractor was 
accurately reproduced with a Erd\"os-R\'enyi network of 300 nodes with a 
mean degree $\overline{\delta} = 6$, each node being made of a difference
equation.\cite{Pat17} The model so-obtained corresponds to an accurate global
model of the dynamics. Notably, this model was constructed from the
measurements of all the variables of the Lorenz system. The main advantage of
such a large model is its flexibility, that is, its ability to capture various dynamical regimes, but it has the disadvantage that the space in which it is 
working is not clearly defined and has a very large dimension ($d_{\rm r}
> 300$ in the work discussed above).

The DDA approach uses a kind of mixed model between the differential model
(\ref{glomod}) and the difference equation (\ref{diffeq}), the left member of 
the latter being replaced with the left member of the former. It is therefore
based on the delay differential equation
\begin{equation}
\label{GeneralDDE}
	\dot{X}  = \mathcal{F}_X 
  = \sum\limits_{i=1}^{N}{\displaystyle  a_i \, \varphi_i (X_{\tau_j})} \,  
\end{equation}
where $X = X_1$ designates the measured variable and $X_{\tau_j}$ some 
delay coordinates. The purpose is not to construct a 
global model reproducing accurately the dynamics but only an 
approximated model for detecting dynamical changes (nonstationarity) or 
classifying different dynamical regimes.\cite{Lai00,Lai13,Lai17} We therefore use a rough model with very 
few terms ($N \leq 3$). Such sparsity in the model prevents 
overparametrization, that is, spurious dynamics induced by overly 
complex models.\cite{Agu95b}
Indeed, delay differential equations are known to already produce complex 
dynamics with only two terms.\cite{Mac77,Far82} Many characteristics of the 
measured dynamics can be captured with two or three terms and appropriate time 
delays.\cite{Lai15c} Based on previous 
works,\cite{Lai00,Lai13,Lai15c,Lai17} it is assumed that these 
characteristics are sufficient to distinguish different dynamical regimes. This DDA model (\ref{GeneralDDE}) is 
a differential equation whose state space is
spanned by delay coordinates $X_{\tau_j}$.

Model (\ref{GeneralDDE}) has two sets of parameters, the fixed parameters 
$\tau_j$ (set during the structure selection) and the free parameters $a_i$
(estimated independently from each data window).
The structure of model (\ref{GeneralDDE}) as well as the delays are determined for each time series.
Then, the free coefficients $a_i$ are determined for each window of the recorded time series.
The data in each window$\left\{ X_1 \right\}$ is normalized to have zero mean and 
unit variance to remove amplitude information before estimating the free 
parameters $a_i$ by using a singular value decomposition (SVD). The 
least-square error
\begin{equation}
  \label{error}
	\rho_X = \sqrt{ \frac{1}{K} \sum_{k=1}^K
	\displaystyle 
	\left( \displaystyle \dot{X} (k) - \mathcal{F}_X (k) \right)^2} \, 
\end{equation}
between the derivatives returned by the DDA model and the derivatives 
computed from the measured time series quantifies the ability of the model to
capture the underlying dynamics. It is known that there is a relationship 
between the model quality and observability.\cite{Let98b,Let02,Let06c}
The signal derivative $\dot{X}_1$ is computed using a five-point center 
derivative.\cite{Mil04} In this work, structure selection (i.e.\ choosing the model form of Eq.~(\ref{GeneralDDE}) and the fixed parameters $\tau_j$)
was performed via an exhaustive 
search over all possible 
three-term models (three monomials: $N=3$)
with 
two delays such that $\tau_j \in [m+1; 60] \delta_t$, where $m=5$ is equal to the number of points for 
estimating the derivative and $\delta_t$ is the sampling time.
 Function $\mathcal{F}$ is made of three monomials 
selected from the possible 
candidates
\begin{equation}
  \begin{array}{rl}
	  \varphi_i \in &
  \left\{ \displaystyle
    X_{\tau_1},
    X_{\tau_2},
    X_{\tau_1}^2,
    X_{\tau_1} \, X_{\tau_2},
	  X_{\tau_2}^2, \right. \\[0.1cm]
	  & \left. \displaystyle ~~~~~~~~~
    X_{\tau_1}^3,
    X_{\tau_1}^2 \, X_{\tau_2},
    X_{\tau_1} \, X_{\tau_2}^2,
    X_{\tau_2}^3
  \right\} \, . 
  \end{array}
\end{equation}
Monomials and delays are selected in an exhaustive search over all possible model forms, i.e. 44, and delay combinations under the restrictions specified above.
Each model is thus characterized by the set of ``fixed'' parameters 
$(\tau_1,\tau_2)$, the corresponding monomials $\varphi_i$, and the 
free parameters $a_i$ which are estimated for each time 
window of the measuredndata. The 
structure providing the model with the lowest $\rho_X$ is retained 
to assess observability according to the model error $\rho_X$. 

As used with reservoir computing,\cite{Car18} the error $\rho_X$ between
the model and the measured data provides a measure of how the system dynamics
may be reconstructed from these data. Indeed, to obtain a reliable 
deterministic model, it is necessary to distinguish every different
state of the system for retrieving the underlying causality. Since the error
is used as a relative measure, it is only needed to have a sufficiently 
flexible functional form for the model as observed with reservoir computing
or with a delay differential equation.
Consequently the smaller the error $\rho_X$, 
the higher the observability provided by the variable $X$. 
This results from previous works where it was shown that the complexity
of the model to approximate was correlated to the observability: the better the
observability provided by the measured variable, the simpler the model to approximate.\cite{Let02,Let06c}
The error $\rho_X$ from the best DDA model is 
computed with an increasing noise amplitude. For each three-dimensional system
and each signal-to-noise ratio (no noise, 20, 10 and 0~dB: where 0dB indicates the variance of the noise matches the variance of the signal), the error $\rho_X$ 
was computed over several hundred pseudoperiods for each time series.

\section{Dynamical systems and observability coefficients }
\label{resu}

\subsection{Low-dimensional systems}

The governing equations of the systems here investigated are reported in Table
\ref{syseq}. 
The symbolic observability coefficients (SOC) and 
the model error $\rho_X$ are reported for each variable of every system in 
Table \ref{syseq}. Parameter values are reported in Table\ \ref{paraval}.

  \begin{table}[htbp]
    \centering
  \caption{Governing equations of each system for which the symbolic 
observability coefficients (SOC) $\eta_{s^3}$ and $\rho_X$ between the DDA model and the measured data with no noise 
contamination are reported. The SOC for variable $x$ of the Hindmarsh-Rose (HR) system 
is corrected as discussed in the main text.For the Chua system, $f(x)=bx+\frac{1}{2}(a-b)(|x+1|-|x-1|)$}
  \label{syseq}
    \begin{tabular}{cllc}
    \\[-0.3cm]
	  \hline \hline
    \\[-0.3cm]
	    System & Equations & SOC & Error \\[0.1cm] \hline
	    R\"ossler 76 & $\dot{x} = -y -z$ & $0.84$ & 0.037
	    \\[0.1cm]
	    \cite{Ros76c}    & $\dot{y} = x +a y$ & 1.0 & 0.022
	    \\[0.1cm]
	    & $\dot{z} = b + z (x -c)$ & $0.56$ & 0.106
	    \\[0.1cm] \hline \\[-0.3cm]
	    R\"ossler 77 & $\dot{x} = -ax -y(1-x^2)$ & 0.56 & $0.0009$
	    \\[0.1cm]
	    \cite{Ros77c} & $\dot{y} = \mu(bx+y-cz)$ & 0.84 & $0.0005$
	    \\[0.1cm]
	    & $\dot{z} = \mu(x+cy-dz)$ & 0.68 & $0.0007$ 
	    \\[0.1cm] \hline \\[-0.3cm]
	    Lorenz 63 & $\dot{x} = \sigma (y -x)$ & 0.78 & 0.02
	    \\[0.1cm]
	    \cite{Lor63} & $\dot{y} = Rx -y -xz$ & 0.36 & 0.039
	    \\[0.1cm]
	    & $\dot{z} = -bz + xy$ & 0.36 & 0.071
	    \\[0.1cm] \hline \\[-0.3cm]
	    Lorenz 84 & $\dot{x} = -y^2 -z^2 -ax + aF$ & 0.36 & 0.061
	    \\[0.1cm]
	    \cite{Lor84} & $\dot{y} = xy -bxz -y + G$ & 0.36 & 0.205
	    \\[0.1cm]
	    & $\dot{z} = bxy +xz -z$ & 0.36 & 0.204
	    \\[0.1cm] \hline \\[-0.3cm]
	    Cord & $\dot{x} = -y -z -ax + aF$ & 0.68 & 0.108
	    \\[0.1cm]
	    \cite{Let12} & $\dot{y} = xy -bxz -y + G$ & 0.36 & 0.198
	    \\[0.1cm]
	    & $\dot{z} = bxy +xz -z$ & 0.36 & 0.232
	    \\[0.1cm] \hline \\[-0.3cm]
	    HR  & $\dot{x} = y-ax^3+bx^2+I-z$ & 0.68 & 0.025
	    \\[0.1cm]
	    \cite{Hin84} & $\dot{y} = c-dx^2-y$ & 0.56 & 0.023
	    \\[0.1cm]
	    & $\dot{z} = r[s(x-x_{\rm R})-z]$ & 1.00 & 0.002
	    \\[0.1cm] \hline \\[-0.3cm]
	    Fisher & $\dot{x} = y$ & 1.00 & 0.003 
	    \\[0.1cm]
	    \cite{Fis99} & $\dot{y} = -ax-by-z$ & 0.84 & 0.004
	    \\[0.1cm]
	    & $\dot{z} = b + x - |x|$ & 0.56 & 0.027
	    \\[0.1cm] \hline \\[-0.3cm]
	    Chua & $\dot{x} = \alpha (-x +y -f(x))$ & 1.00 & 0.05 
	    \\[0.1cm]
	    \cite{Chu86} & $\dot{y} = x-y+z$ & 0.84 & 0.068
	    \\[0.1cm]
	    & $\dot{z} = - \beta y $ & 1.00 & 0.066
	    \\[0.1cm] \hline \\[-0.3cm]
	    Duffing & $\dot{x} = y$ & 1.00 & 0.022 
	    \\[0.1cm]
	    \cite{Duf18,Men00} & $\dot{y} = - \mu y +x -x^3 + u$ & 0.86 & 0.08
	    \\[0.1cm]
	    & $\dot{u} = v$ & 0.00 & 0.00 \\[0.1cm]
	    & $\dot{v} = - \omega^2 u$ & 0.00 & 0.00 
	    \\[0.1cm] \hline \\[-0.3cm]
	    R\"ossler 79 & $\dot{x} = -y-z$ & 0.75 & 0.005 
	    \\[0.1cm]
	    \cite{Ros79a} & $\dot{y} = x + ay + w$ & 0.83 & 0.001
	    \\[0.1cm]
	    & $\dot{z} = b + xz$ & 0.44 & 0.079 \\[0.1cm]
	    & $\dot{w} = - cz + dw$ & 0.63 & 0.006 
	    \\[0.1cm] \hline \\[-0.3cm]
	    H\'enon-Heiles & $\dot{x} = u$ & 0.64 & 0.0005
	    \\[0.1cm]
	    \cite{Hen64} & $\dot{y} = v$ & 0.64 & 0.0004
	    \\[0.1cm]
	    & $\dot{u} = -x -2xy$ & 0.44 & 0.0009 \\[0.1cm]
	    & $\dot{v} = -y -y^2 -x^2$ & 0.44 & 0.0008 \\[0.1cm]
	  \hline \hline
    \end{tabular}
  \end{table}

\begin{figure}[ht]
  \centering
	\includegraphics[width=0.48\textwidth]{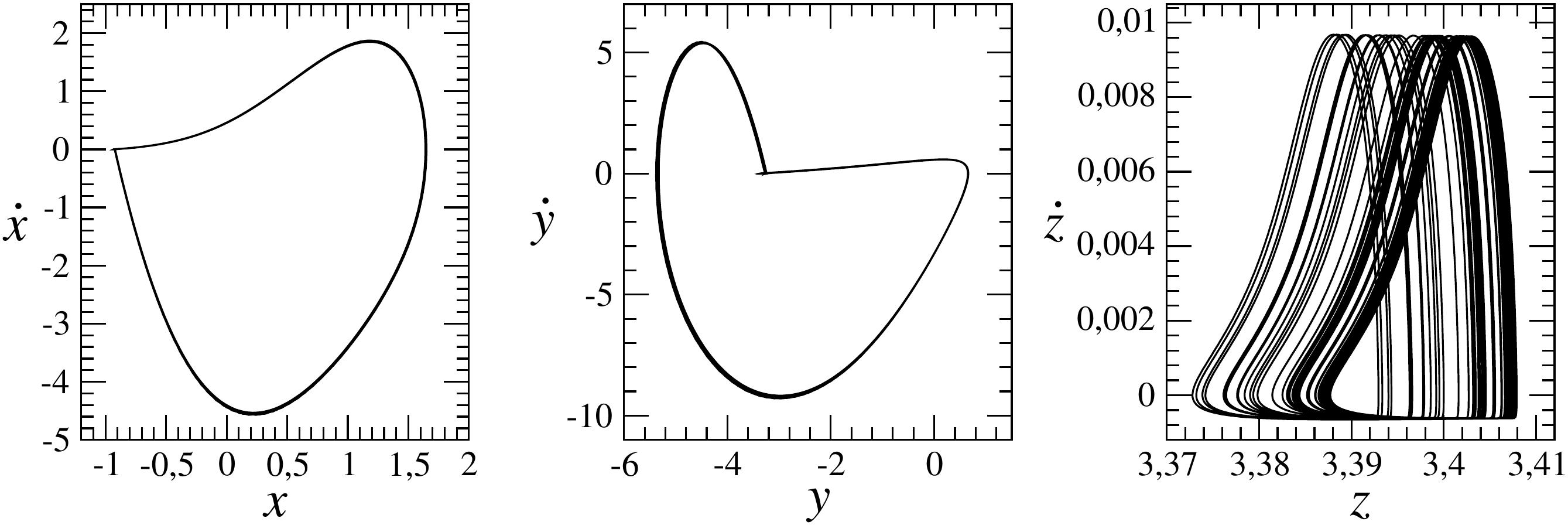} \\[-0.3cm]
  \caption{Differential embedding induced by each of the three variables of the
	Hindmarsh-Rose system.}
	\label{HRdiffemb}
\end{figure}

The R\"ossler 76\cite{Ros76c}, Lorenz 84\cite{Lor84},
Cord\cite{Let12}, Hindmarsh-Rose\cite{Hin84} (HR) and Fisher\cite{Fis99} systems have 
no symmetry. The Hindmarsh-Rose system is known to be problematic when variable 
$x$ or $z$ is measured, for two different reasons.\cite{Agu17} When variable 
$z$ is measured, the observability matrix
\begin{equation}
  {\cal O}_z =
  \left[
    \begin{array}{ccc}
      0 & 0 & 1 \\[0.1cm]
      rs & 0 & -r \\[0.1cm]
	    rs \left( \displaystyle x \chi -r \right) & rs & r(r-s) 
    \end{array}
  \right]
\end{equation}
where $\chi = 2b - 3ax$ becomes singular when $r$ is too small (Det${\cal O}_z
= r^2 s^2$): the observability can be null for $r = 0$ and full for $r \neq 0$
(this is also true for $s$, but $s$ is commonly significantly different from 
0). When variable $x$ is measured, although the observability matrix 
${\cal O}_x$ is never singular (Det ${\cal O}_x = r-1$), the plane projection
of the differential embedding induced by variable $x$ does not reveal the
chaotic nature of the underlying dynamics, contrary to what is clearly 
provided by variable $z$ (Fig.\ \ref{HRdiffemb}). As discussed by Aguirre {\it
et al},\cite{Agu17} the observability matrix 
\begin{equation}
  {\cal O}_x =
  \left[
    \begin{array}{ccc}
      1 & 0 & 0 \\[0.1cm]
      \chi x & 1 & 1 \\[0.1cm]
      O_{31}^x & \chi x -1 & - \chi x +r 
    \end{array}
  \right]
\end{equation}
where
\begin{equation}
  \begin{array}{rl}
	  O_{31}^x = & \displaystyle \chi^2 x^2 -rs - 2bx + 2(b-3a) \\ 
	  & ~~~ \times
	\left[ \left( \displaystyle I + x^2 (b-ax)+y -z \right) \right]
  \end{array}
\end{equation}
has a determinant Det~${\cal O}_x$ whose polynomial nature is cancelled by
the contributions of $O_{32}$ and $O_{33}$ but this is not structurally 
stable. Any perturbation in one of these two elements would lead to a 
determinant vanishing for a subset of the state space. This is not detected
by the symbolic observability coefficients. If we keep the polynomial nature
of elements $O_{32}$ and $O_{33}$, the symbolic observability matrix would be
\begin{equation}
  {\cal O}_x =
  \left[
    \begin{array}{ccc}
      1 & 0 & 0 \\[0.1cm]
      \1 & 1 & 1 \\[0.1cm]
      \1 & \1 & \1 
    \end{array}
  \right] \, .
\end{equation}
The corresponding corrected symbolic observability coefficient is thus
$\eta'_{x^3} = 0.68$. The corrected ranking of variables is therefore 
$z \vartriangleright x \vartriangleright y$. This ranking will be used in the 
subsequent analysis.

\begin{table}[ht]
  \centering
  \caption{Parameter values of the investigated systems.}
  \label{paraval}
  \begin{tabular}{ccccccccc}
    \\[-0.3cm]
	  \hline \hline
    \\[-0.3cm]
	  R\"ossler 76 & $a=0.52$ & $b=2$ & $c=4$ \\[0.1cm]
	  \hline \\[-0.3cm]
	  R\"ossler 77 & $a = 0.03$ & $b=0.3$ & $c=2$ & $d=0.5$ \\
	  & $\mu = 0.1$
	  \\[0.1cm]
	  \hline \\[-0.3cm]
	  Lorenz 63 & $\sigma = 10$ & $b = 8/3$ & $R=28$ 
	  \\[0.1cm]
	  \hline \\[-0.3cm]
	  Lorenz 84 & $a= 0.28$ & $b=4$ & $F= 8$ & $G=1$
	  \\[0.1cm]
	  \hline \\[-0.3cm]
	  Cord & $a= 0.28$ & $b=4$ & $F= 8$ & $G=1$
	  \\[0.1cm]
	  \hline \\[-0.3cm]
	  HR & $a=1$ & $b=3$ & $c=1$ & $d=5$ \\
      & $I = 3.29$
	  & \multicolumn{2}{l}{$x_{\rm R}=\frac{8}{5}$}
	  \\[0.1cm]
	  \hline \\[-0.3cm]
	  Fisher & $a = 0.3$ & $b = 0.097$ & 
	  \\[0.1cm]
	  \hline \\[-0.3cm]
	  Chua & $\alpha = 9$ & $\beta = \frac{100}{7}$ & $a = - \frac{8}{7}$  &  \\ & $b = - \frac{5}{7}$
	  \\[0.1cm]
	  \hline \\[-0.3cm]
	  Duffing & $\mu = 0.3$ & $\omega = 1.2$ \\ & $x_0 = 1$ & $y_0 = 0$ 
	  & $u_0 = 0.5$ & $v_0 = 0$
	  \\[0.1cm]
	  \hline \\[-0.3cm]
	  R\"ossler 79 & $a = 0.25$ & $b = 3$ & $c = 0.5$ & $d = 0.05$ 
	  \\[0.1cm]
	  \hline \hline
  \end{tabular}
\end{table}

The other systems have symmetry properties as follows.
The Lorenz 63 system\cite{Lor63} is equivariant under a ${\cal R}_z$ rotation symmetry 
around the $z$-axis.\cite{Let94a,Let01} Variables $x$ and $y$ are mapped into 
their opposite ($-x$ and $-y$, respectively) while variable $z$ is invariant 
under the rotation symmetry. At least two variables must be measured to 
correctly reconstruct the rotation symmetry.\cite{Kin92} The R\"ossler 77\cite{Ros77c} , Chua 
circuit\cite{Chu86} and the driven
Duffing systems\cite{Duf18,Men00} are equivariant under an inversion symmetry. Such a symmetry
can be recovered from a single variable and, consequently, should not blur 
the observability analysis. The driven Duffing system is in fact a 
four-dimensional system, a conservative harmonic oscillator driving the 
dissipative Duffing oscillator: it is thus a semi-dissipative (or 
semi-conservative) system.\cite{Men00} When variable $u$ (or $v$) is recorded , a 
periodic orbit is obtained while variable $x$ (or $y$) provides a chaotic 
state portrait. Since a chaotic driving signal necessarily implies a chaotic 
response, it is obvious that $u$ drives $x$ and not the opposite. It can 
therefore be concluded, without further analysis, that the system is not
observable from $u$ (or $v$). Thus, we only have to determine the 
observability from variable $x$ and $y$, respectively. The Fisher system and 
the Chua circuit have a piecewise nonlinearity. They will be useful to test
whether DDA is robust against discontinuous nonlinearity. 

All these systems but three --- the Lorenz 84, the Cord, and the H\'enon-Heiles\cite{Hen64}
systems --- have at least one variable providing a good observability 
($\eta > 0.75$) of the original state space. The H\'enon-Heiles system is 
conservative and one may guess that the observability problem will be more
sensitive since the invariant domain of the state space has a dimension close
to 3, and not 2 as for all the other systems which are strongly dissipative.

\subsection{A higher-dimensional system}

\begin{figure}[ht]
  \centering
  \begin{tabular}{cc}
    \includegraphics[width=0.25\textwidth]{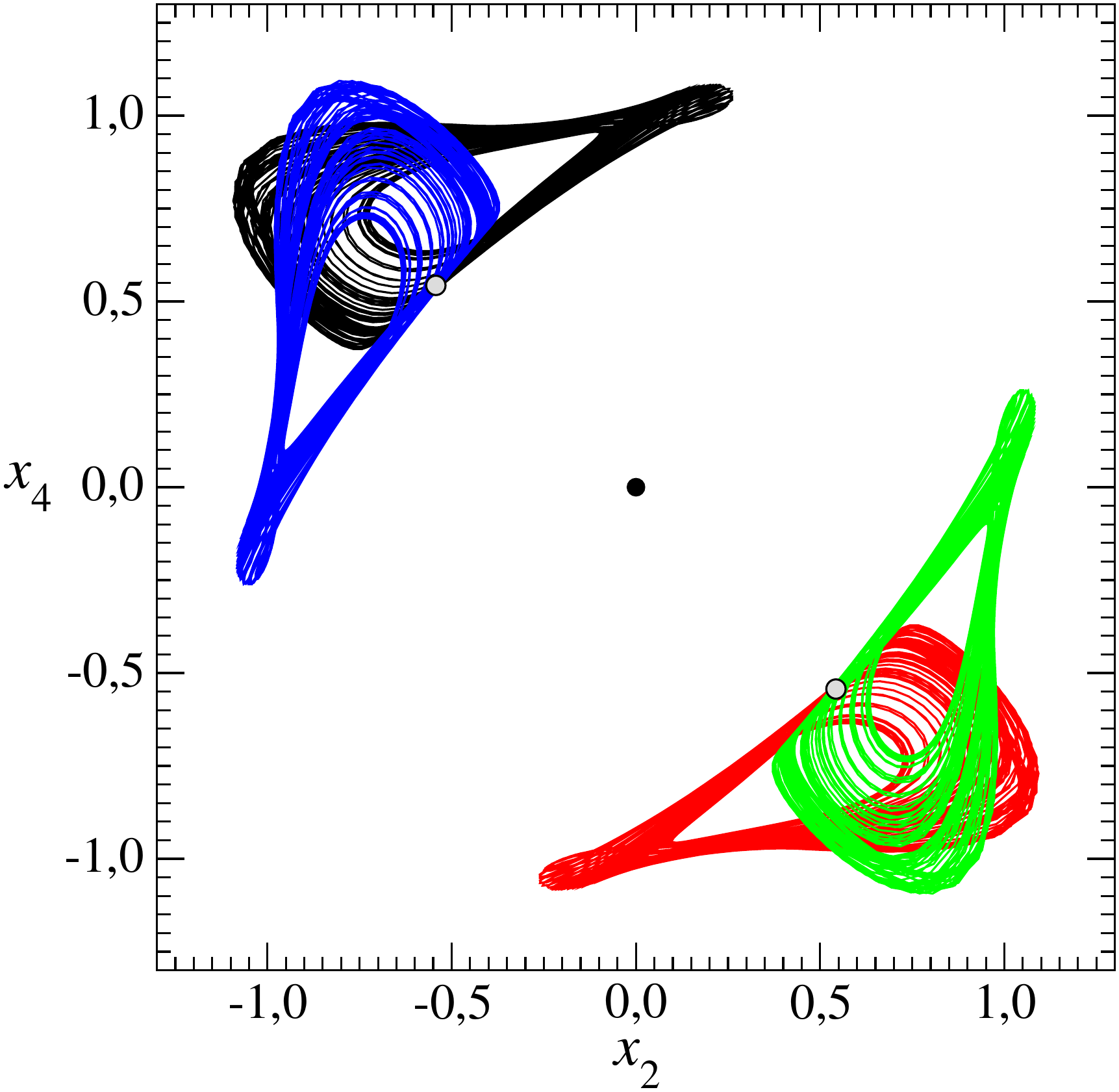} &
    \includegraphics[width=0.22\textwidth]{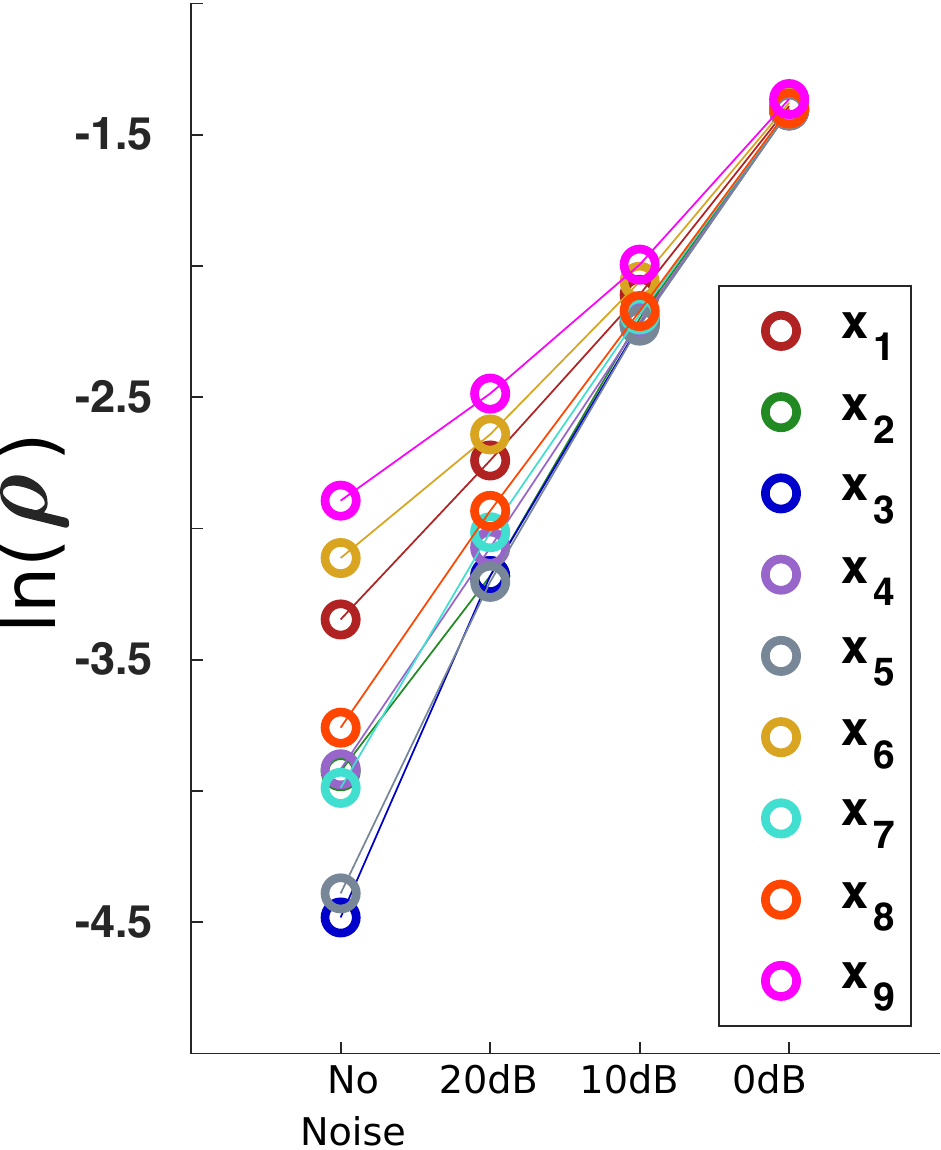} \\
	  \multicolumn{2}{c}{(a)} \\[0.2cm]
    \includegraphics[width=0.22\textwidth]{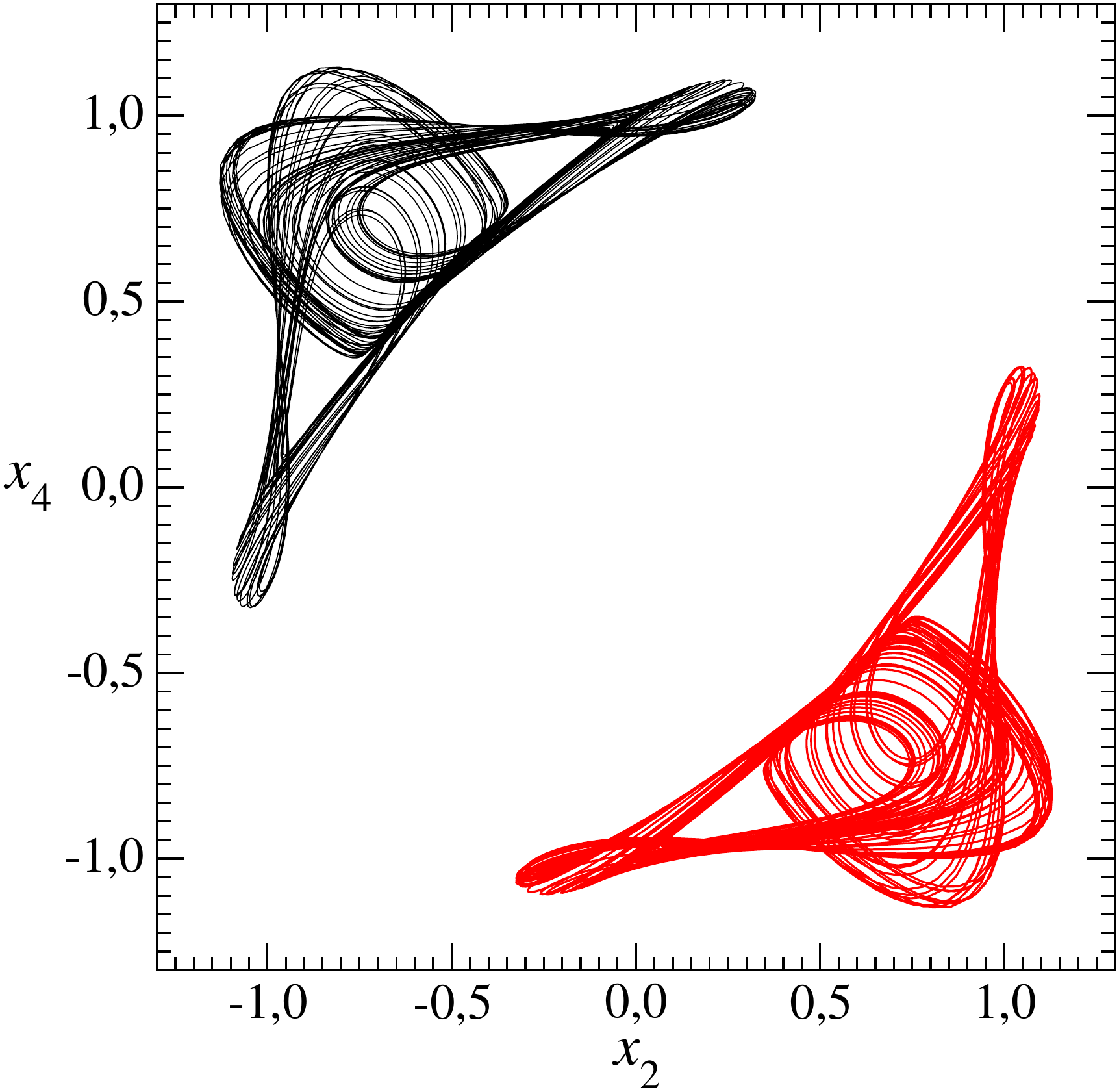} &
    \includegraphics[width=0.22\textwidth]{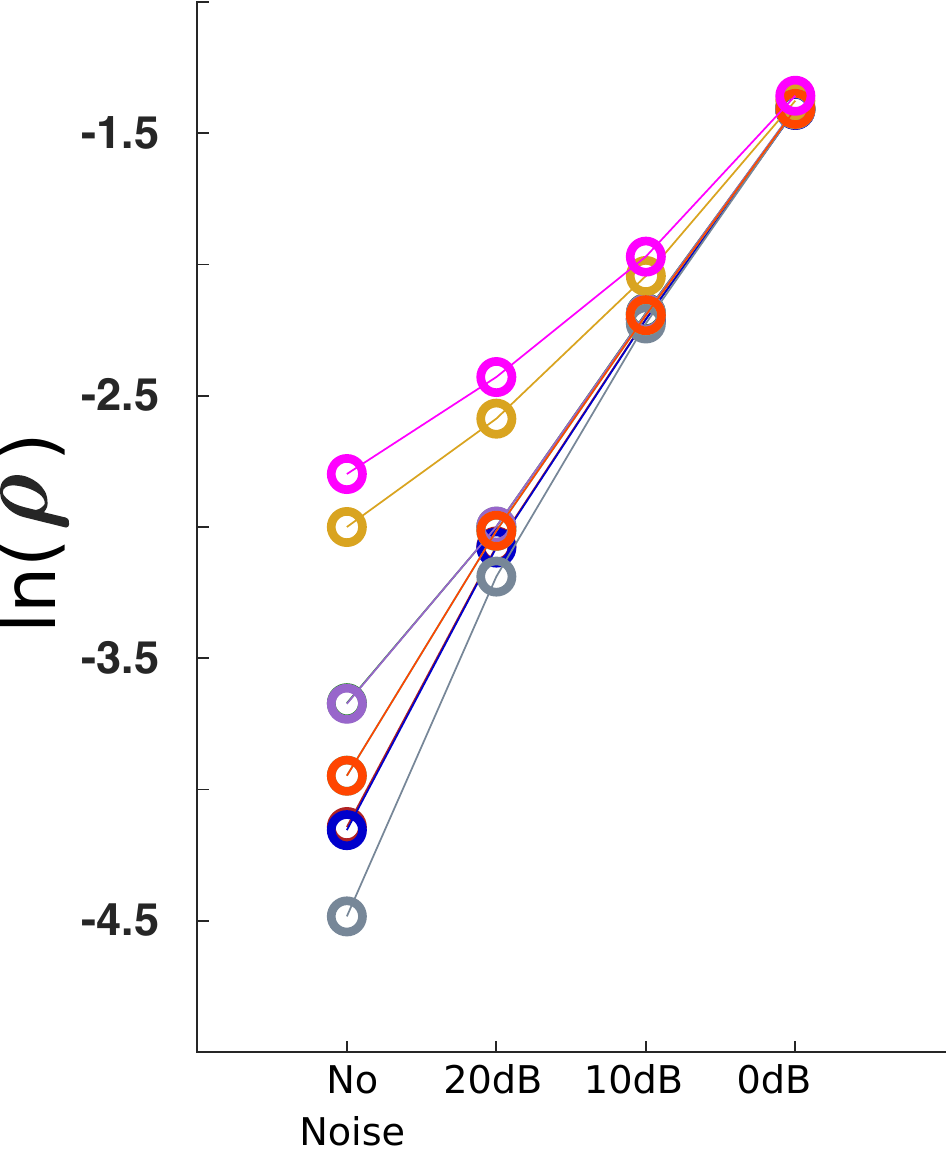} \\
	  \multicolumn{2}{c}{(b) } \\[0.2cm]
    \includegraphics[width=0.22\textwidth]{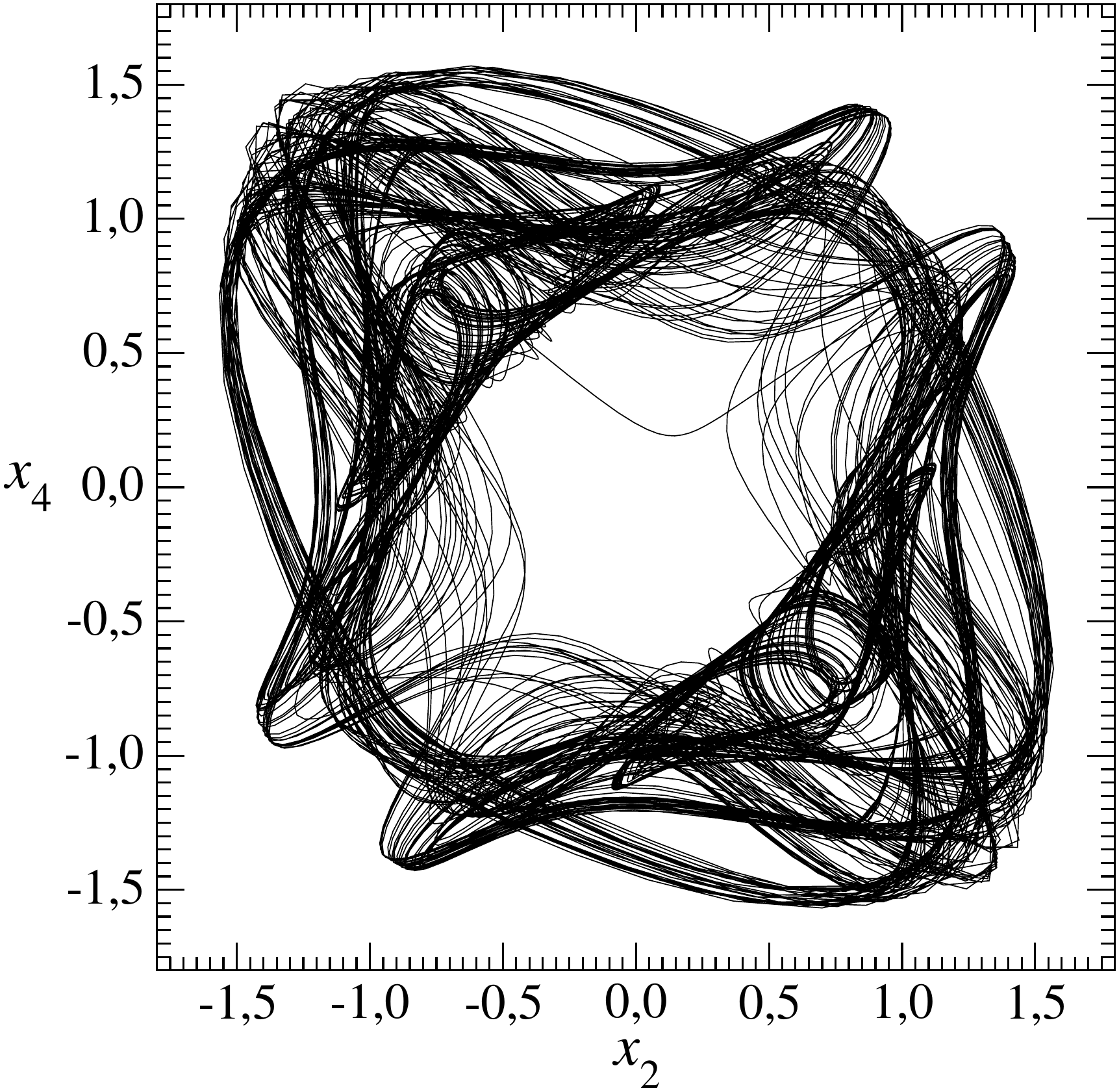} &
    \includegraphics[width=0.22\textwidth]{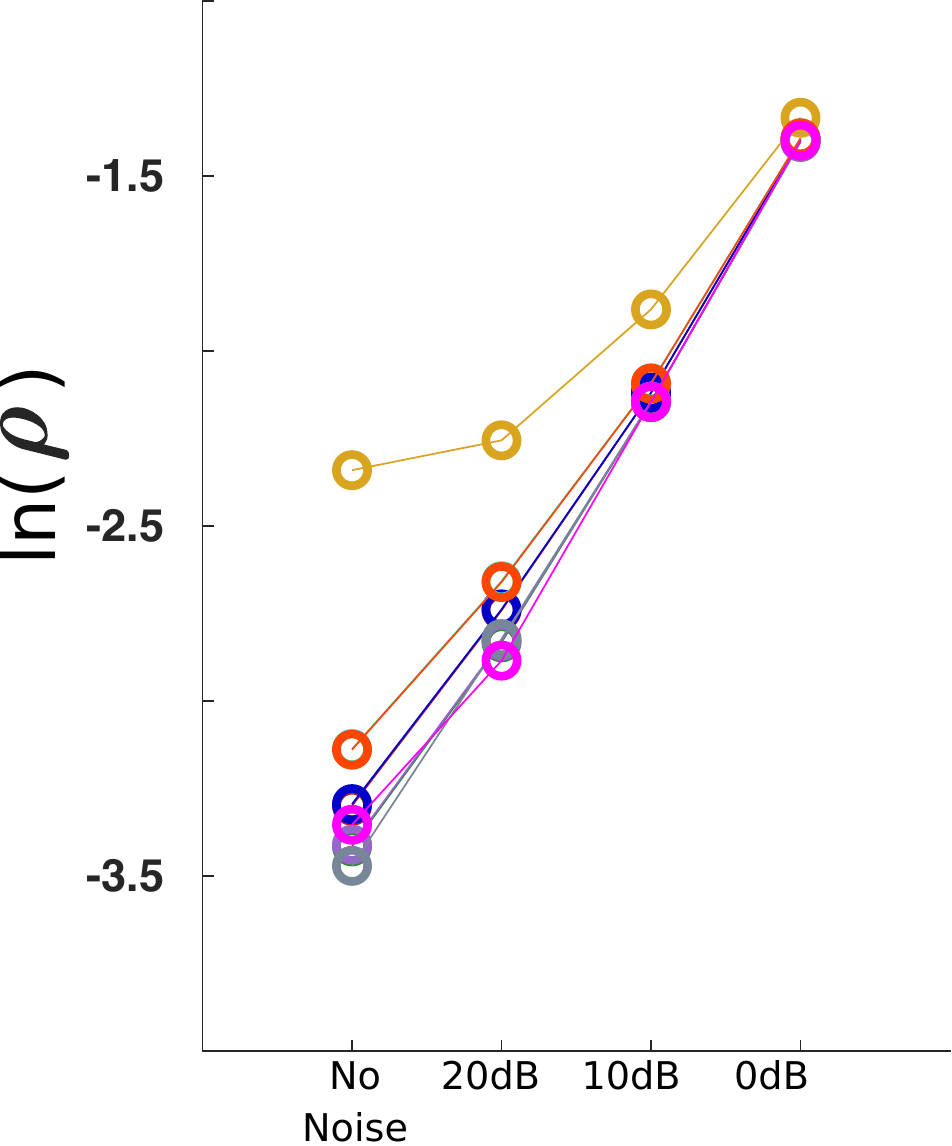} \\
	  \multicolumn{2}{c}{(c) } \\
    \includegraphics[width=0.22\textwidth]{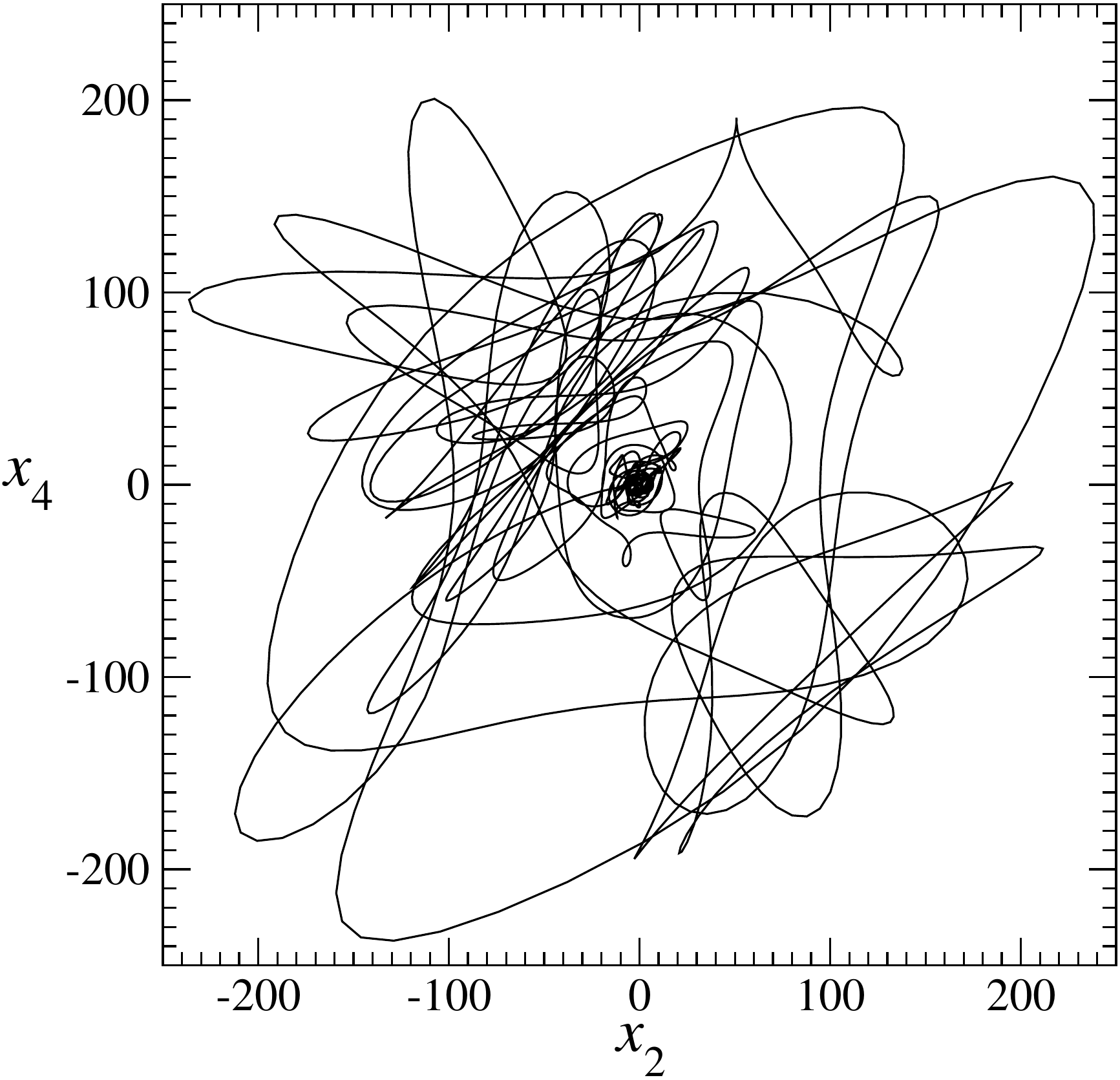} &
    \includegraphics[width=0.22\textwidth]{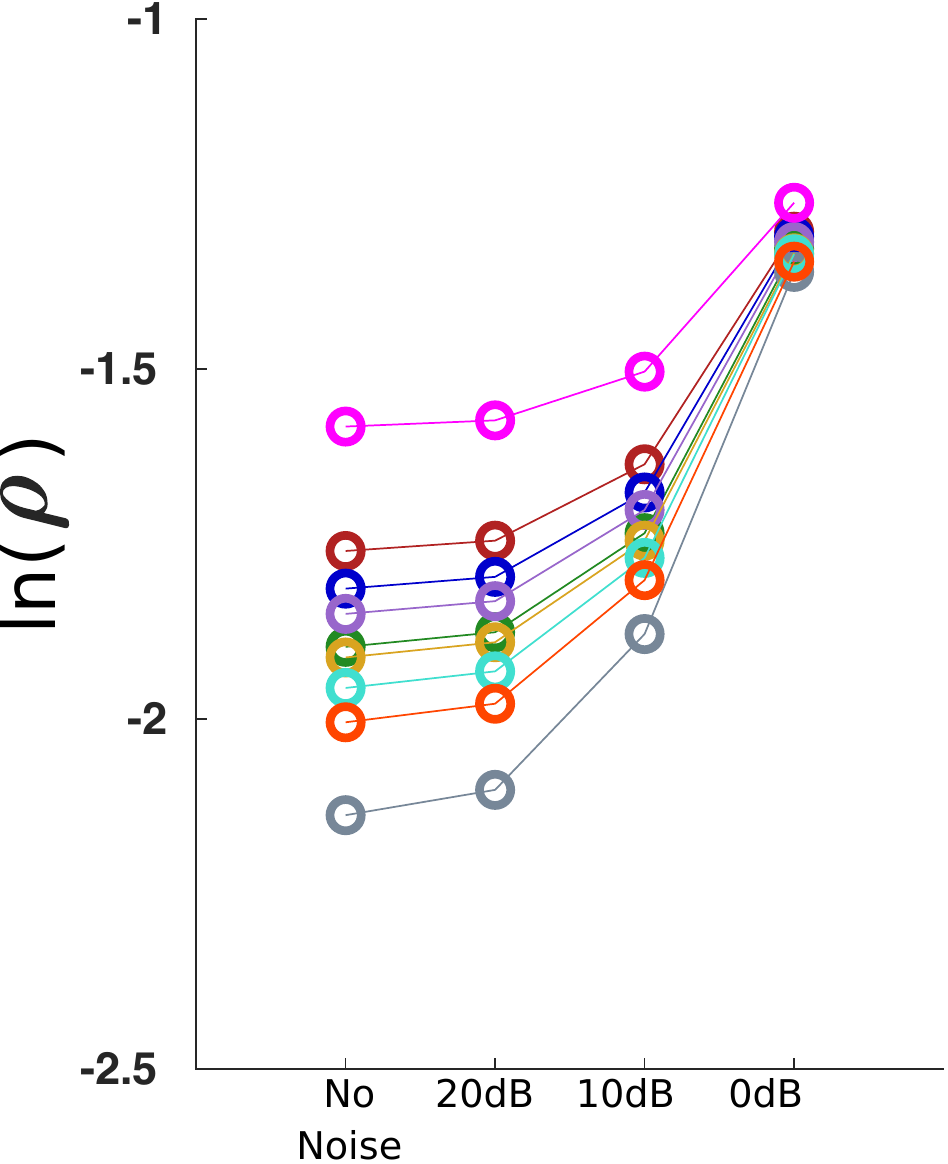} 
	  \\
	  \multicolumn{2}{c}{(d)} \\[-0.3cm]
  \end{tabular}
  \caption{Chaotic attractor produced by the 9D Lorenz system
    (\ref{9Dsym}). (a) $R = 14.22$, (b) $R = 14.30$, (c) $R = 15.10$, (d) $R = 45.00$.
Other parameter values: $a=0.5$, and $\sigma=0.5$. 
When there are co-existing attractors, they are plotted in different colors in the plane projections of the state space.
}
  \label{9DLorenz}
\end{figure}

The Lorenz 63 system results from a Galerkin expansion of the
Navier-Stokes equations for Rayleigh-B\'enard convection.\cite{Sal62} It is 
also possible to have a higher-dimensional expansion in retaining more Fourier 
components. One of them lead to the 9D Lorenz system\cite{Rei98}
\begin{equation}
  \label{9DLOR}
  \left\{
    \begin{array}{l}
	    \dot{x_{1}} = -\sigma (b_1 x_1 + b_2 x_7) + x_4 (b_4 x_4 -x_2)
	    + b_3x_3x_5 \\[0.1cm]
	    \displaystyle
    \dot{x_{2}} = -\sigma{x_2} + x_1x_4 -x_2x_5 +x_4x_5 - \frac{\sigma x_9}{2} 
	    \\[0.1cm]
      \dot{x_{3}} = \sigma (b_2 x_8 -b_1x_3) +x_2x_4-b_4{x_2^2}-b_3{x_1x_5}
	    \\[0.1cm]
	    \displaystyle
    \dot{x_{4}} = -\sigma{x_4}-x_2x_3-x_2x_5+x_4x_5 + \frac{\sigma x_9}{2} 
	    \\[0.1cm]
	    \displaystyle
    \dot{x_{5}} = -\sigma{b_5}x_5 + \frac{x_2^2}{2}- \frac{x_4^2}{2} \\[0.1cm]
      \dot{x_{6}} = -b_6x_6+x_2x_9-x_4x_9 \\[0.1cm]
      \dot{x_{7}} = -b_1{x_7}-Rx_1+2x_5x_8-x_4x_9 \\[0.1cm]
      \dot{x_{8}} = -b_1{x_8}+Rx_3-2x_5x_7+x_2x_9 \\[0.1cm]
	    \dot{x_{9}} = -x_9 + (R +2 x_6) (x_4-x_2) +x_4x_7-x_2x_8 
    \end{array}
  \right.
\end{equation}
where 
\begin{equation}
  \label{9Dlor_bs}
  \left\{
    \begin{array}{ll}
	    \displaystyle
	    b_1 =  4\frac{1+a^2}{1+2a^2} &
	    \displaystyle
     b_2 =  \frac{1+2a^2}{2(1+a^2)} \\[0.3cm]
	    \displaystyle
	    b_3 =  2\frac{1-a^2}{1+a^2} &
	    \displaystyle
     b_4 =  \frac{a^2}{1+a^2} \\[0.3cm]
	    \displaystyle
	    b_5 =  \frac{8a^2}{1+2a^2} &
	    \displaystyle
     b_6 =  \frac{4}{1+2a^2} 
    \end{array}
  \right.
\end{equation}
This 9D Lorenz system is equivariant.\cite{Gil07} Depending on the $R$-values, 
the attractor produced may be asymmetric [Fig.\ \ref{9DLorenz}(a)] or 
symmetric [Fig.\ \ref{9DLorenz}(b)]. The symbolic observability coefficients 
are
\begin{equation}
    \label{9Dsym}
  \left\{
    \begin{array}{l}
	    \displaystyle
    \eta_{x_1^9} = \eta_{x_3^9} = \eta_{x_7^9} = \eta_{x_8^9} = 0.04 \\[0.1cm]
	    \displaystyle
    \eta_{x_2^9} = \eta_{x_4^9} =  0.03 \\[0.1cm]
	    \displaystyle
    \eta_{x_5^9} = \eta_{x_6^9} = \eta_{x_9^9} =  0 
    \end{array}
  \right.
\end{equation}
leading to
\[ x_1 = x_3 = x_7 = x_8 \vartriangleright x_2 = x_4 \vartriangleright x_5 = 
x_6 = x_9 \]
Notice that every variable offers an extremely poor observability of the 
original state space. It was shown that, at least five variables need to be 
measured for having a good observability ($\eta > 0.75$) of the original state 
space.\cite{Let18} Moreover, for a sufficiently large $R$-value ($R=45$), the 
behavior is hyperchaotic. One of the characteristics of this highly developed
behavior is that there are two different time scales. We will therefore 
investigate whether the observability assessed with DDA is dependent on parameter values, that is, on bifurcation affecting the 
symmetry properties (order-4 or order-2 asymmetric chaos, symmetric chaos
and hyperchaos). 

\section{DDA ranking}
\label{DDAran}

\begin{figure*}[htbp]
  \centering
  \begin{tabular}{ccccc}
	  \includegraphics[width=0.232\textwidth]{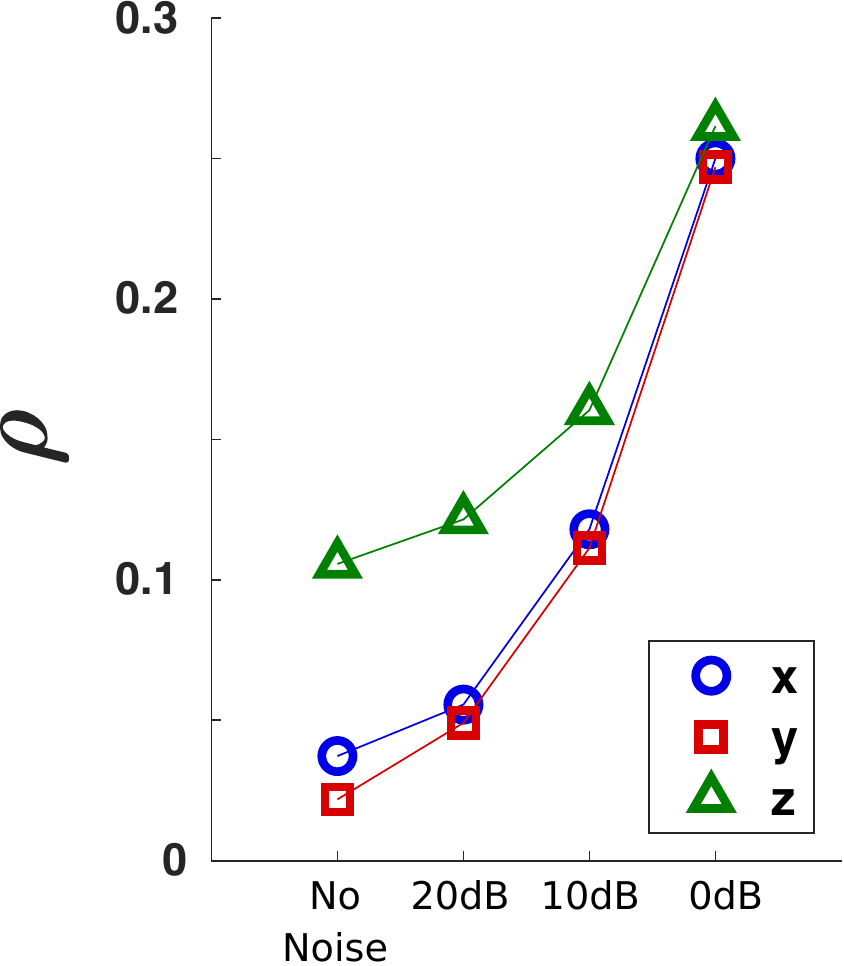}  & ~~~ &
	  \includegraphics[width=0.232\textwidth]{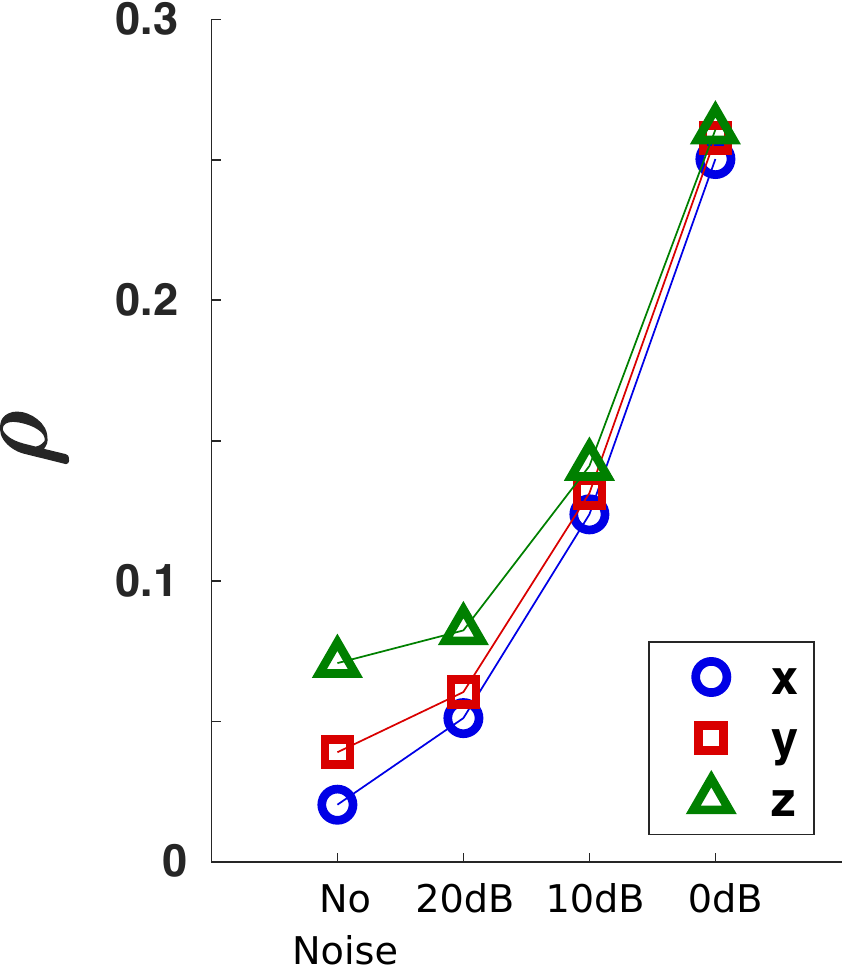}  & ~~~ &
	  \includegraphics[width=0.232\textwidth]{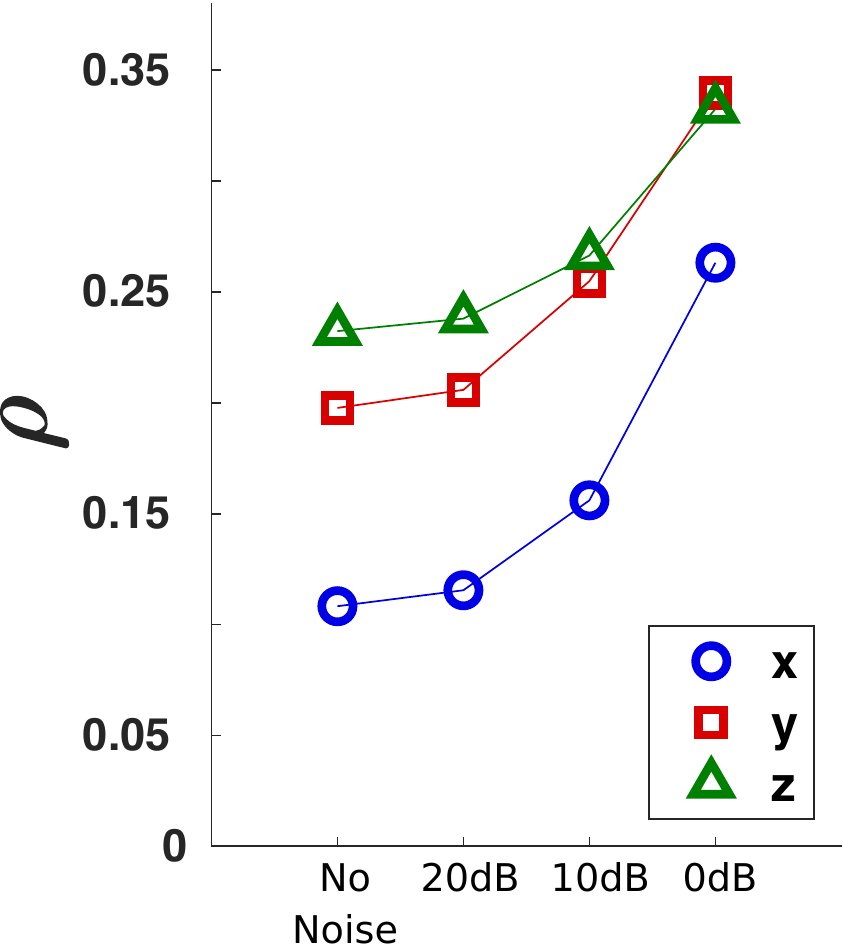}  \\
	  {\footnotesize (a) }  &  &
	  {\footnotesize (b) } & &
	  {\footnotesize (c) } \\[0.2cm]
	  \includegraphics[width=0.232\textwidth]{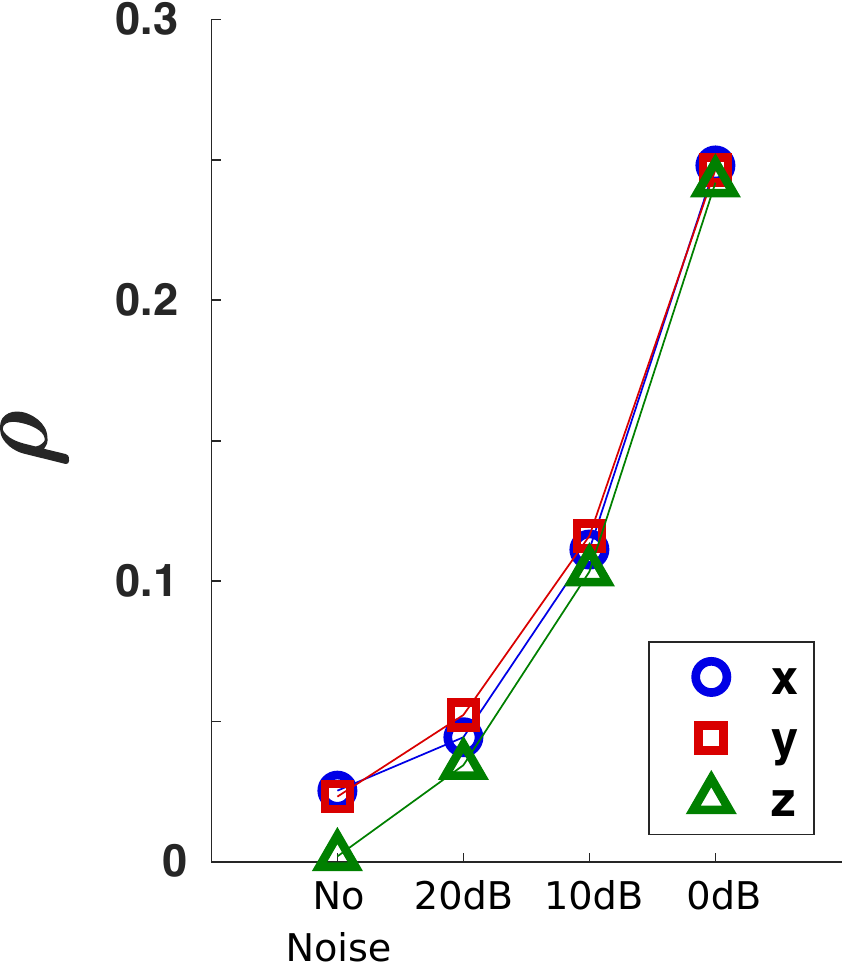}  & ~~~ &
	  \includegraphics[width=0.35\textwidth]{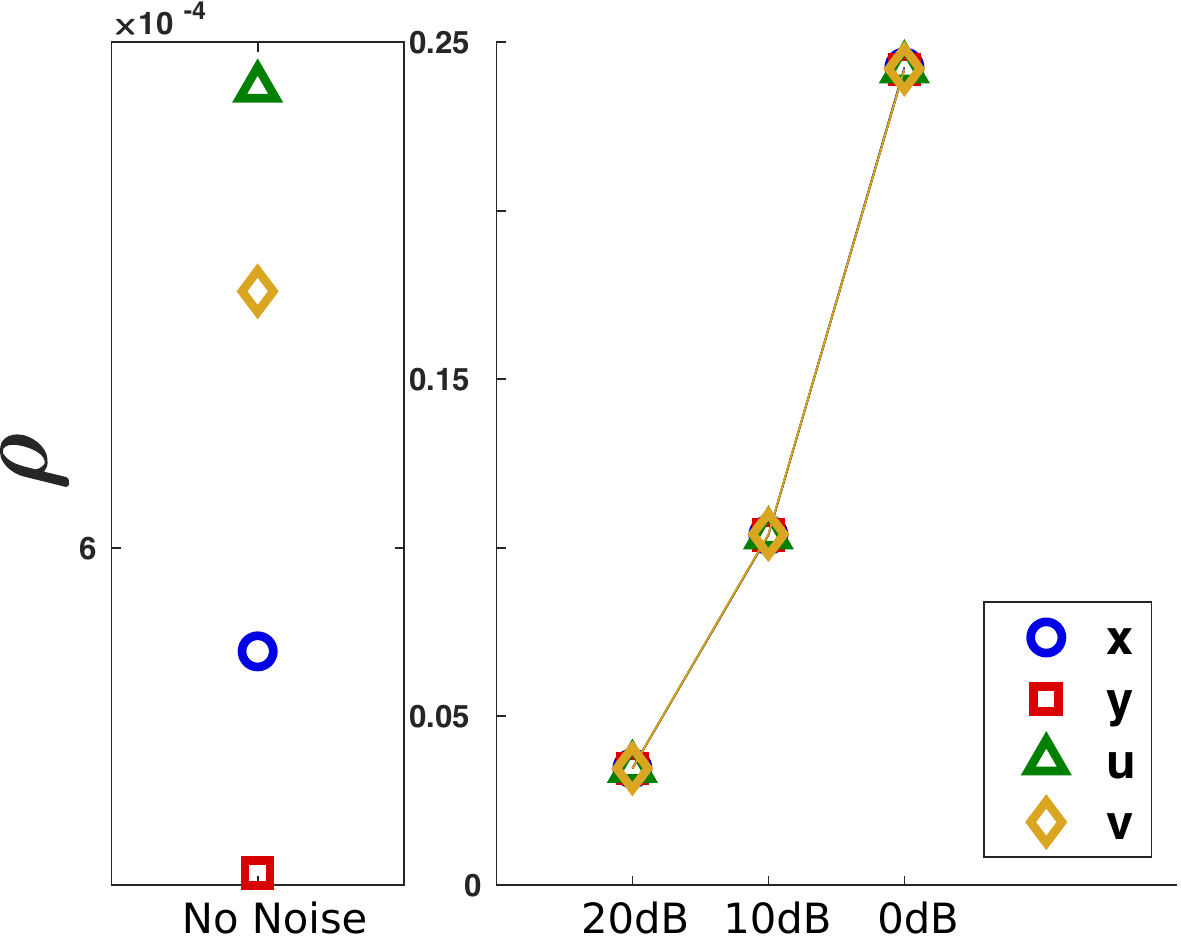}  & ~~~ &
	  \includegraphics[width=0.232\textwidth]{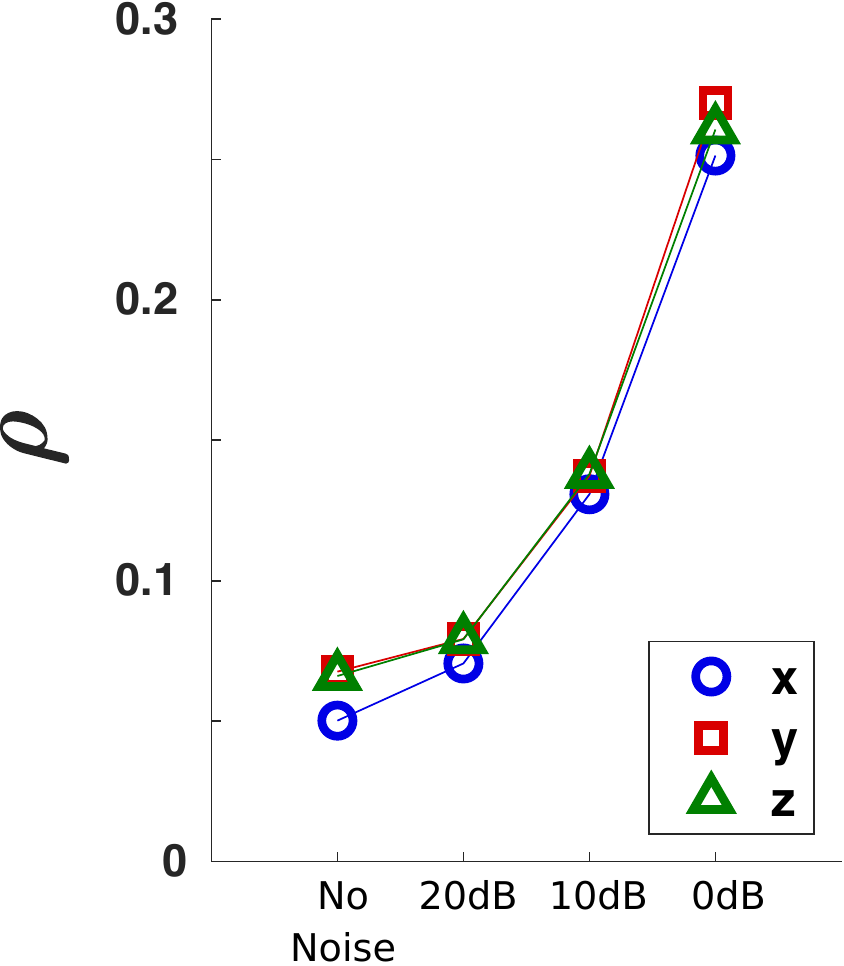}  \\
	  {\footnotesize (d) } & &
	  {\footnotesize (e) } & &
	  {\footnotesize (f) }
	  \\
  \end{tabular}
  \caption{Error $\rho_X$ versus a decreasing 
signal-to-noise ratio for some of the different systems investigated in this paper. (a) R\"ossler 76 system, (b) Lorenz 63 system, (c) Cord system, (d) Hindmarsh-Rose system, (e) H\'enon-Heiles system, (f) Chua system.}
  \label{DDerrors}
\end{figure*}

The structure of the best DDA models ${\cal F}_X$ are reported in Appendix
\ref{modform}, Table\ \ref{funform} 
along with the corresponding time delays retained for identifying the
free parameters. As examples, $\rho_X$ for some systems with
increasing noise are shown in Figs.~\ref{DDerrors}. For no noise, $\rho_X$
is reported in Table\ \ref{syseq}.

The rankings for variables according to increasing symbolic observability
coefficients (SOC), decreasing $\rho_X$  for DDA, and when available in the literature, for decreasing reservoir
computing (RC) and singular value decomposition observability 
(SVDO) are summarized in Table\ \ref{compar} for all low dimensional 
systems ($d \leq 4$). The results for the R\"ossler 76, R\"ossler 77, Fisher, driven Duffing and R\"ossler 79\cite{Ros79a} systems are in a perfect agreement with
the SOC. The discontinuity of the Fisher system does not perturb the analysis.
The hyperchaotic nature of the R\"ossler 79 system was not problematic for 
correctly assessing observability. 

The Lorenz 63, Lorenz 84, Cord and Hindmarsh-Rose systems show close agreement between DDA and 
SOC. For the Lorenz 63 system, variable $x$ was 
correctly detected as providing the best observability but variable $z$ was 
found to offer worse observability than variable $y$, a feature which is not
predicted by the SOC due to a problem inherent to the symmetry
involved. For the Cord system, while no single variable provides good observability for the original state space, DDA correctly ranks $x$ as providing the best observability. However, DDA ranks $z$ as providing worse observability than variable $y$, while SOC ranks them with equivalent observability.
For the Hindmarsh-Rose system, variable $z$ provides full
observability and is associated with the lowest $\rho_X$. However,
there is some discrepancy between DDA and SOC since, as assessed with
DDA, $y$ provides a slightly higher observability than $x$. Results for the H\'enon-Heiles system are quite equivalent to
the SOC.Variables $x$ and $y$ are more observable than $u$ and $v$, however, $x$($u$) is more observable than $y$($v$) instead of showing equivalent observability.

For the Chua circuit, the variable $x$ contains a piecewise nonlinearity and has full observability, and DDA correctly ranks $x$ as the most observable. DDA also ranks variable
$y$ with the worst observability, which is in agreement with SOC. However, variable $z$ has only slightly better observability than $y$, whereas it should be equivalent to $x$.

\begin{table*}[htbp]
  \centering
  \caption{Ranking variables according to the observability as assessed by
the symbolic observability coefficients (SOC), DDA analysis, reservoir computing (RC) and singular value decomposition observability
(SVDO). A perfect agreement with the SOC is indicated by a $\bullet$. When 
the variable providing the best observability is correctly detected or when 
$=$ is replaced with $\thickapprox$ or $\vartriangleright$, a $\circ$
is reported.}
  \label{compar}
  \begin{tabular}{cccccc}
	  \\[-0.3cm]
    \hline \hline
	  \\[-0.2cm]
	  System & SOC & DDA & RC & SVDO \\[0.1cm]
 \hline
	  \\[-0.3cm]
	 R\"ossler 76
	  & $y \vartriangleright x \vartriangleright z$ 
	  & $\bullet$ & $x \vartriangleright y \vartriangleright z$ 
	  & $\bullet$ 
	  \\[0.1cm]
	  \hline \\[-0.3cm]
	  R\"ossler 77
	  & $y \vartriangleright z \vartriangleright x$ 
	  & $\bullet$  & --- & --- \\[0.1cm]
	  \hline \\[-0.3cm]
	  Lorenz 63
	  & $x \vartriangleright y = z$ 
	  & $\circ$ 
	  & $y \vartriangleright x \vartriangleright z$ 
	  & $\circ$ \\[0.1cm]
	  \hline \\[-0.3cm]
	   Lorenz 84
	  & $x = y = z$ 
	  & $\circ$ & --- & $\bullet$ \\[0.1cm]
	  \hline \\[-0.3cm]
	  Cord
	  & $x \vartriangleright y = z$ 
	  & $\circ$ & --- & $\circ$ \\[0.1cm]
	  \hline \\[-0.3cm]
      Hindmarsh-Rose
	  & $z \vartriangleright x \vartriangleright y$
	  & $\circ$ & --- & $\bullet$  \\[0.1cm]
	  \hline \\[-0.3cm]
	Fisher
	  & $x \vartriangleright y \vartriangleright z$ 
	  & $\bullet$ & --- & --- \\[0.1cm]
	  \hline \\[-0.3cm]
	Chua
	  & $x = z \vartriangleright y$ 
	  & $\circ$
	  & $\bullet$ & $\circ$ \\[0.1cm]
	  \hline \\[-0.3cm]
	  Duffing & $x \vartriangleright y \vartriangleright u = v$ 
	  & $\bullet$ & --- & --- \\[0.1cm]
	  \hline \\[-0.3cm]
	 R\"ossler 79 
	  & $y \vartriangleright x \vartriangleright w \vartriangleright z$ 
	  & $\bullet$ & $x \vartriangleright y \vartriangleright z \vartriangleright w$ 
	  & $x \vartriangleright y \vartriangleright w \vartriangleright z$ 
	  \\[0.1cm]
	  \hline \\[-0.3cm]
	 H\'enon-Heiles 
	  & $x = y \vartriangleright u = v$
	  & $\circ$ 
	  & $\circ$  
	  & --- 
	  \\[0.1cm]
    \hline \hline
  \end{tabular}
\end{table*}

When compared to the two other data-based techniques, DDA performs better than
RC for the R\"ossler 76, R\"ossler 79 and the Lorenz 63 systems but not for the Chua
circuit. Compared to the SVDO, the DDA approach provides equivalent results
for all the systems investigated by these two techniques, but does perform better for the hyperchaotic R\"ossler 79 system in correctly identifying the variable $y$ as providing the best observability, a feature which  missed by the SVDO.

For most of the systems, these results are robust against noise contamination, 
at least up to a signal-to-noise ratio greater than 10 dB: below this ratio, 
results can be blurred and observability can no longer be reliably assessed
using DDA. A similar robustness was observed with SVDO. It was not 
investigated with RC.

Note that another interesting data-based technique for assessing observability
was proposed by Parlitz and co-workers.\cite{Par14b} It was only tested with 
the R\"ossler 76 system (and the H\'enon map, not investigated here). It would
be interesting to further investigate its performance but this is out of the 
scope of this paper.

The results for the 9D Lorenz system are not so clear. The first reason is that
this system is nearly unobservable from a single variable. The SOC are 
nearly saturated (close to 0) with nonlinear elements as revealed by the 
symbolic Jacobian matrix of the 9D Lorenz system (\ref{9DLOR}), namely
\begin{equation}
  {\cal J}^{\rm sym} =
  \left[
    \begin{matrix}
	    1 & \1 & \1 & \1 & \1 &  0 &  1 &  0 &  0 \\ 
	   \1 & \1 &  0 & \1 & \1 &  0 &  0 &  0 &  1 \\ 
	   \1 & \1 &  1 & \1 & \1 &  0 &  0 &  1 &  0 \\ 
	    0 & \1 & \1 & \1 & \1 &  0 &  0 &  0 &  1 \\ 
	    0 & \1 &  0 & \1 & \1 &  0 &  0 &  0 &  0 \\ 
	    0 & \1 &  0 & \1 &  0 &  1 &  0 &  0 & \1 \\ 
	    1 &  0 &  0 & \1 & \1 &  0 &  1 & \1 & \1 \\ 
	    0 & \1 &  1 &  0 & \1 &  0 & \1 &  1 &  0 \\ 
	    0 & \1 & \1 &  0 & \1 & \1 & \1 & \1 &  1 \\ 
    \end{matrix}
  \right] \, , 
\end{equation}
which illustrates most of the couplings between variables are nonlinear. 
Considering only the observability provided by a single variable is here investigated, and that the SOC are all close to 0, one may conclude that the 9D Lorenz system is not observable from a single variable.

Results provided by DDA are shown in Fig.\ \ref{9DLorenz} where it is seen
that variables cannot be easily ranked, particularly when $R$ is increased. 
Results are summarized in Table\ \ref{obs9DLor} as follows. For 
each $R$-values, the rankings of the variables are reported --- from 1 for the 
variable offering the best observability to 9 for the one providing the poorest
observability --- and compared to the ranking provided by the SOC. The results
are strongly dependent on $R$-value in a way which does not allow to 
extract a clear tendency. Variable $x_5$ with a null observability as assessed
by the SOC (and analytically) is found to provide the best observability as 
assessed by DDA. Nevertheless, this is in agreement with the successful
three-dimensional global model obtained from this variable for 
$R = 14.22$,\cite{Rei98} that is, at least for this $R$-value, the dynamics can
be correctly reconstructed for recovering the underlying  determinism.

\begin{table}[ht]
  \centering
  \caption{Observability of the 9D Lorenz system as assessed with the symbolic
	observability coefficients (SOC) and DDA.}
	\label{obs9DLor}
  \begin{tabular}{ccccccccccc}
	  \\[-0.3cm]
	  \hline \hline
	  \\[-0.3cm]
  & $R$ & $x_1$ & $x_2$ & $x_3$ & $x_4$ & $x_5$ & $x_6$ & $x_7$ & $x_8$ & $x_9$ 
	  \\[0.1cm]
	  \hline
	  \\[-0.3cm]
    SOC & --- & 1 & 2 & 1 & 2 & 3 & 3 & 1 & 1 & 3 \\[0.1cm]
	  \hline
	  \\[-0.3cm]
    DDA & 14.22 & 7 & 4 & 1 & 5 & 2 & 8 & 3 & 6 & 9 \\
	& 14.30 & 3 & 7 & 2 & 6 & 1 & 8 & 5 & 4 & 9 \\
	& 15.10 & 5 & 2 & 6 & 3 & 1 & 9 & 8 & 7 & 4 \\
	  & 45.00 & 8 & 5 & 7 & 6 & 1 & 4 & 3 & 2 & 9 \\[0.1cm]
	  \hline \hline
  \end{tabular}
\end{table}

It should be pointed out that looking for full observability (i.e. being
able to ``reconstruct'' each of the non-measured variables) is not the same 
thing as looking for an embedding. Especially for large $d$-dimensional 
systems producing an attractor which can be embedded within a space whose dimension $d_{\rm R}$ is lower than the dimension $d$ of the original state space. Full observability ensures 
the existence of an embedding, the opposite is not necessarily true. Here 
DDA selects the variable which provides the best reconstructed space. If 
compared with the results provided by the SOC with multivariate 
measurements,\cite{Let18} variables $x_2$, $x_4$, $x_5$, and $x_6$ are always
among the six variables selected for providing a full observability. DDA 
returns three of them as providing the best observability, $x_2$, $x_4$, and 
$x_5$ (Table\ \ref{obs9DLor}). Variable $x_6$, the single one which is 
invariant under the symmetry of this system, is identified as a variable 
providing a poor observability. Once again, symmetry induces difficulties for assessing observability.

\section{Conclusion}
\label{conc}

The ability to infer the state of a system from a scalar output depends on 
which system variable is measured. We have introduced a numerical approach 
using the error between a DDA model and measured data to assess the 
observability provided by the measured variables in several chaotic systems. We 
compared these measures with symbolic observability coefficients, which are 
determined directly from the system's equations. Our measure overall reliably ranks variables according to the observability they 
provide about the original state space. The largest discrepancy was 
obtained for a large-dimensional (9D Lorenz) system. The smaller the model error, the better the observability 
provided. The assessment of observability is quite robust against noise 
contamination in the majority of the systems here considered.

There are two situations in which our approaches may face some 
complications. The first one is a common one. Inconsistencies in assessing 
observability are known for systems with symmetry properties, particularly with 
variables left invariant. The second one is also a typical one: when the 
dimension of the system increases, the observability of the state space provided
by a single variable becomes very poor and assessing observability is delicate. 
Our approach is thus very reliable for low-dimensional systems without symmetry properties, even with a 
signal-to-noise ratio as commonly encountered in experiments.

As in most of the other techniques, variables of different systems cannot be 
compared to each other. This is a common limitation in assessing observability 
that is only overcome by using an analytical approach, such as by computing 
explicitly the observability matrix or by using the symbolic observability coefficients. A
kind of normalization should be considered to have, for instance, the error
$\rho_y$ of variable $y$ of the R\"ossler 76 system (which has full observability) smaller than for variable 
$y$ of the R\"ossler 77 system. This problem is more challenging than it may 
appear. It was, for instance, never solved for the observability coefficients
computed along a trajectory using a relationship extracted from the system's
equations or using SVD applied to a reconstructed space.

\acknowledgments
C.~Letellier wishes to thank Irene Sendi\~na-Nadal for her assistance in 
computing the symbolic observability coefficients for the 9D Lorenz system. This work was supported by the National Institute of Health
(NIH)/NIBIB (Grant No. R01EB026899-01) and by the National Science Foundation Graduate Research Fellowship (Grant No. DGE-1650112).

\subsection*{Data Availability}

The data that support the findings of this study are available from the 
corresponding author upon reasonable request.

\appendix

\section{Functional form of DDA models}
\label{modform}

\begin{table}[htbp]
  \centering
  \caption{Functional forms of the DDA models for each variable of the systems 
investigated. The time delays are expressed in terms of $\delta_t$, the 
sampling time at which variable $X$ is recorded.}
  \label{funform}
  \begin{tabular}{ccccccc}
	  \hline \hline
	  \\[-0.3cm]
	  & & $a_1$ & $a_2$ & $a_3$ & $\tau_1$ & $\tau_2$ 
	  \\[0.1cm] \hline
	  \\[-0.3cm]
	  R\"ossler 76 & $\mathcal{F}_x$ & $X_{\tau_1}$ & $X_{\tau_2}$ 
	  & $X_{\tau_1}^3$ & $6 \, \delta_t$ & $7 \, \delta_t$ 
	  \\[0.1cm]
	  & $\mathcal{F}_y$ & $X_{\tau_1}$ & $X_{\tau_2}$ 
	  & $X_{\tau_1}^3$ & $6 \, \delta_t$ & $7 \, \delta_t$ \\[0.1cm]
	  & $\mathcal{F}_z$ & $X_{\tau_1}$ & $X_{\tau_2}$ 
	  & $X_{\tau_1}^3$ & $6 \, \delta_t$ & $7 \, \delta_t$ 
	  \\[0.1cm] \hline
	  \\[-0.3cm]
	  R\"ossler 77 & $\mathcal{F}_x$ & $X_{\tau_1}$ & $X_{\tau_2}$ 
	  & $X_{\tau_1}^3$ & $6 \, \delta_t$ & $7 \, \delta_t$ \\[0.1cm]
	  & $\mathcal{F}_y$ & $X_{\tau_1}$ & $X_{\tau_2}$ 
	  & $X_{\tau_1}^3$ & $7 \, \delta_t$ & $6 \, \delta_t$ \\[0.1cm]
	  & $\mathcal{F}_z$ & $X_{\tau_1}$ & $X_{\tau_2}$ 
	  & $X_{\tau_1}^3$ & $6 \, \delta_t$ & $7 \, \delta_t$ 
	  \\[0.1cm] \hline
	  \\[-0.3cm]
	  Lorenz 63 & $\mathcal{F}_x$ & $X_{\tau_1}$ & $X_{\tau_2}$ 
	  & $X_{\tau_1}^3$ & $6 \, \delta_t$ & $7 \, \delta_t$ \\[0.1cm]
	  & $\mathcal{F}_y$ & $X_{\tau_1}$ & $X_{\tau_1}^3$
	  & $X_{\tau_1} X_{\tau_2}^2$ & $6 \, \delta_t$ & $19 \, \delta_t$
	  \\[0.1cm]
	  & $\mathcal{F}_z$ & $X_{\tau_1}$ & $X_{\tau_1}^2$ 
	  & $X_{\tau_2}^2$ & $18 \, \delta_t$ & $6 \, \delta_t$ 
	  \\[0.1cm] \hline
	  \\[-0.3cm]
	  Lorenz 84 & $\mathcal{F}_x$ & $X_{\tau_1}$ & $X_{\tau_2}$ 
	  & $X_{\tau_1}^2$ & $7 \, \delta_t$ & $6 \, \delta_t$ \\[0.1cm]
	  & $\mathcal{F}_y$ & $X_{\tau_1}$ & $X_{\tau_1}^3$ 
	  & $X_{\tau_1} X_{\tau_2}^2$ & $6 \, \delta_t$ & $28 \, \delta_t$ 
	  \\[0.1cm]
	  & $\mathcal{F}_z$ & $X_{\tau_1}$ & $X_{\tau_1} X_{\tau_2}$ 
	  & $X_{\tau_1} X_{\tau_2}^2$ & $6 \, \delta_t$ & $60 \, \delta_t$ 
	  \\[0.1cm] \hline
	  \\[-0.3cm]
	  Cord & $\mathcal{F}_x$ & $X_{\tau_1}$ & $X_{\tau_1}^3$ 
	  & $X_{\tau_1}^2 X_{\tau_2}$ & $7 \, \delta_t$ & $51 \, \delta_t$ 
	  \\[0.1cm]
	  & $\mathcal{F}_y$ & $X_{\tau_1}$ & $X_{\tau_1} X_{\tau_2}$ 
	  & $X_{\tau_1} X_{\tau_2}^2$ & $6 \, \delta_t$ & $18 \, \delta_t$ 
	  \\[0.1cm]
	  & $\mathcal{F}_z$ & $X_{\tau_1}$ & $X_{\tau_2}$ 
	  & $X_{\tau_1}^2$ & $6 \, \delta_t$ & $7 \, \delta_t$ 
	  \\[0.1cm] \hline
	  \\[-0.3cm]
	  HR & $\mathcal{F}_x$ & $X_{\tau_1}$ & $X_{\tau_1} X_{\tau_2}$ 
	  & $X_{\tau_2}^3$ & $6 \, \delta_t$ & $9 \, \delta_t$ 
	  \\[0.1cm]
	  & $\mathcal{F}_y$ & $X_{\tau_1}^2$ & $X_{\tau_1}^2 X_{\tau_2}$ 
	  & $X_{\tau_2}^3$ & $25 \, \delta_t$ & $6 \, \delta_t$ 
	  \\[0.1cm]
	  & $\mathcal{F}_z$ & $X_{\tau_1}$ & $X_{\tau_2}$ 
	  & $X_{\tau_1}^2$ & $6 \, \delta_t$ & $7 \, \delta_t$ 
	  \\[0.1cm] \hline
	  \\[-0.3cm]
	  Fisher & $\mathcal{F}_x$ & $X_{\tau_1}$ & $X_{\tau_2}$ 
	  & $X_{\tau_1}^2$ & $7 \, \delta_t$ & $6 \, \delta_t$ 
	  \\[0.1cm]
	  & $\mathcal{F}_y$ & $X_{\tau_1}$ & $X_{\tau_2}$ 
	  & $X_{\tau_1}^2$ & $6 \, \delta_t$ & $7 \, \delta_t$ 
	  \\[0.1cm]
	  & $\mathcal{F}_z$ & $X_{\tau_1}$ & $X_{\tau_2}$
	  & $X_{\tau_1}^2$ & $7 \, \delta_t$ & $6 \, \delta_t$ 
	  \\[0.1cm] \hline
	  \\[-0.3cm]
	  Chua & $\mathcal{F}_x$ & $X_{\tau_1}$ & $X_{\tau_2}$ 
	  & $X_{\tau_1}^3$ & $6 \, \delta_t$ & $7 \, \delta_t$ 
	  \\[0.1cm]
	  & $\mathcal{F}_y$ & $X_{\tau_1}$ & $X_{\tau_2}$ 
	  & $X_{\tau_1}^3$ & $7 \, \delta_t$ & $6 \, \delta_t$ 
	  \\[0.1cm]
	  & $\mathcal{F}_z$ & $X_{\tau_1}$ & $X_{\tau_2}$ 
	  & $X_{\tau_1}^3$ & $13 \, \delta_t$ & $32 \, \delta_t$ 
	  \\[0.1cm] \hline
	  \\[-0.3cm]
	  Duffing & $\mathcal{F}_x$ & $X_{\tau_1}$ & $X_{\tau_2}$ 
	  & $X_{\tau_1}^3$ & $6 \, \delta_t$ & $7 \, \delta_t$ 
	  \\[0.1cm]
	  & $\mathcal{F}_y$ & $X_{\tau_1}$ & $X_{\tau_2}$ 
	  & $X_{\tau_1}^3$ & $6 \, \delta_t$ & $7 \, \delta_t$ 
	  \\[0.1cm]
	  & $\mathcal{F}_u$ & $X_{\tau_1}$ & $X_{\tau_2}$ 
	  & $X_{\tau_1}^2$ & $38 \, \delta_t$ & $37 \, \delta_t$ 
	  \\[0.1cm] 
	  & $\mathcal{F}_v$ & $X_{\tau_1}$ & $X_{\tau_2}$ 
	  & $X_{\tau_1}^2$ & $38 \, \delta_t$ & $37 \, \delta_t$ 
	  \\[0.1cm] \hline
	  \\[-0.3cm]
	  9D Lorenz & $\mathcal{F}_{1,3,5}$ & $X_{\tau_1}$ & $X_{\tau_2}$ 
	  & $X_{\tau_1}^2$ & $7 \, \delta_t$ & $6 \, \delta_t$ 
	  \\[0.1cm]
	  $R = 14.22$ 
	  & $\mathcal{F}_{4,7,8}$ & $X_{\tau_1}$ & $X_{\tau_2}$ 
	  & $X_{\tau_1}^3$ & $7 \, \delta_t$ & $6 \, \delta_t$ 
	  \\[0.1cm]
	  & $\mathcal{F}_{6,9}$ & $X_{\tau_1}$ & $X_{\tau_2}$ 
	  & $X_{\tau_1}^3$ & $6 \, \delta_t$ & $7 \, \delta_t$ 
	  \\[0.1cm]
	  & $\mathcal{F}_2$ & $X_{\tau_1}$ & $X_{\tau_1} X_{\tau_2}$ 
	  & $X_{\tau_2}^2$ & $47 \, \delta_t$ & $14 \, \delta_t$ 
	  \\[0.1cm] \hline
	  \\[-0.3cm]
	  9D Lorenz & $\mathcal{F}_{1-5,7,8}$ & $X_{\tau_1}$ & $X_{\tau_2}$ 
	  & $X_{\tau_1}^3$ & $6 \, \delta_t$ & $7 \, \delta_t$ 
	  \\[0.1cm]
	  $R = 14.30$ 
	  & $\mathcal{F}_{6,9}$ & $X_{\tau_1}$ & $X_{\tau_2}$ 
	  & $X_{\tau_1}^3$ 
	  & $7 \, \delta_t$ & $6 \, \delta_t$ 
	  \\[0.1cm] \hline
	  \\[-0.3cm]
	  9D Lorenz & $\mathcal{F}_{1,3,7,8}$ & $X_{\tau_1}$ & $X_{\tau_2}$ 
	  & $X_{\tau_1}^2$ & $7 \, \delta_t$ & $6 \, \delta_t$
	  \\[0.1cm]
	  $R = 15.10$ 
	  & $\mathcal{F}_{2,4,5,9}$ & $X_{\tau_1}$ & $X_{\tau_2}$ 
	  & $X_{\tau_1}^3$ & $6 \, \delta_t$ & $7 \, \delta_t$ 
	  \\[0.1cm]
	  & $\mathcal{F}_6$ & $X_{\tau_1}$ & $X_{\tau_1}^2$ 
	  & $X_{\tau_2}^2$ & $25 \, \delta_t$ & $6 \, \delta_t$ 
	  \\[0.1cm] \hline
	  \\[-0.3cm]
	  9D Lorenz & $\mathcal{F}_1$ & $X_{\tau_1}$ & $X_{\tau_1}X_{\tau_2}$ 
	  & $X_{\tau_1}^3$ & $6 \, \delta_t$ & $11 \, \delta_t$
	  \\[0.1cm]
	  $R = 45$ 
	  & $\mathcal{F}_{2}$ & $X_{\tau_1}$ & $X_{\tau_1}^3$ 
	  & $X_{\tau_1}X_{\tau_2}^2$ & $6 \, \delta_t$ & $57 \, \delta_t$ 
	  \\[0.1cm]
	  & $\mathcal{F}_{3}$ & $X_{\tau_1}$ & $X_{\tau_1}^3$ 
	  & $X_{\tau_2}^3$ & $6 \, \delta_t$ & $7 \, \delta_t$ 
	  \\[0.1cm]
	  & $\mathcal{F}_{4}$ & $X_{\tau_1}$ & $X_{\tau_1}^3$ 
	  & $X_{\tau_1}X_{\tau_2}^2$ & $6 \, \delta_t$ & $44 \, \delta_t$ 
	  \\[0.1cm]
	  & $\mathcal{F}_5$ & $X_{\tau_1}^3$ & $X_{\tau_1}^2X_{\tau_2}$ 
	  & $X_{\tau_2}^3$ & $10 \, \delta_t$ & $6 \, \delta_t$ 
	  \\[0.1cm]
	  & $\mathcal{F}_6$ & $X_{\tau_1}^2$ & $X_{\tau_1}X_{\tau_2}$ 
	  & $X_{\tau_1}^3$ & $10 \, \delta_t$ & $23 \, \delta_t$ 
	   \\[0.1cm]
	  & $\mathcal{F}_7$ & $X_{\tau_1}$ & $X_{\tau_1}X_{\tau_2}$ 
	  & $X_{\tau_2}^3$ & $6 \, \delta_t$ & $9 \, \delta_t$ 
	  \\[0.1cm]
	  & $\mathcal{F}_8$ & $X_{\tau_1}$ & $X_{\tau_1}X_{\tau_2}$ 
	  & $X_{\tau_2}^3$ & $6 \, \delta_t$ & $10 \, \delta_t$ 
	  \\[0.1cm]
	  & $\mathcal{F}_9$ & $X_{\tau_1}$ & $X_{\tau_1}^2X_{\tau_2}$ 
	  & $X_{\tau_2}^3$ & $6 \, \delta_t$ & $9 \, \delta_t$ 	  \\[0.1cm]
	  \hline \hline
	  \end{tabular}
\end{table}

\addtocounter{table}{-1}
\begin{table}[htbp]
  \centering
  \caption{Cont}
  \label{funform2}
  \begin{tabular}{ccccccc}
	  \hline \hline
	  \\[-0.3cm]
	  & & $a_1$ & $a_2$ & $a_3$ & $\tau_1$ & $\tau_2$ 
	  \\[0.1cm] \hline
	  \\[-0.3cm]
	  R\"ossler 79 & $\mathcal{F}_x$ & $X_{\tau_1}$ & $X_{\tau_2}$ 
	  & $X_{\tau_1}^2$ & $6 \, \delta_t$ & $7 \, \delta_t$ 
	  \\[0.1cm]
	  & $\mathcal{F}_y$ & $X_{\tau_1}$ & $X_{\tau_2}$ 
	  & $X_{\tau_1}^3$ & $7 \, \delta_t$ & $6 \, \delta_t$ 
	  \\[0.1cm]
	  & $\mathcal{F}_z$ & $X_{\tau_1}$ & $X_{\tau_2}$ 
	  & $X_{\tau_1}^2$ & $6 \, \delta_t$ & $7 \, \delta_t$ 
	  \\[0.1cm] 
	  & $\mathcal{F}_w$ & $X_{\tau_1}$ & $X_{\tau_2}$ 
	  & $X_{\tau_1}^3$ & $7 \, \delta_t$ & $6 \, \delta_t$ 
	  \\[0.1cm] \hline
	  \\[-0.3cm]
	  H\'enon-Heiles & $\mathcal{F}_x$ & $X_{\tau_1}$ & $X_{\tau_2}$ 
	  & $X_{\tau_1}^3$ & $6 \, \delta_t$ & $7 \, \delta_t$ 
	  \\[0.1cm]
	  & $\mathcal{F}_y$ & $X_{\tau_1}$ & $X_{\tau_2}$ 
	  & $X_{\tau_1}^2$ & $6 \, \delta_t$ & $7 \, \delta_t$ 
	  \\[0.1cm]
	  & $\mathcal{F}_u$ & $X_{\tau_1}$ & $X_{\tau_2}$ 
	  & $X_{\tau_1}^3$ & $6 \, \delta_t$ & $7 \, \delta_t$ 
	  \\[0.1cm] 
	  & $\mathcal{F}_v$ & $X_{\tau_1}$ & $X_{\tau_2}$ 
	  & $X_{\tau_1}^3$ & $7 \, \delta_t$ & $6 \, \delta_t$ 
	  \\[0.1cm]
	  \hline \hline
	  \end{tabular}
\end{table}

%


\end{document}